%% file: Doug0127.tex
\documentclass[twocolumn]{aastex6}

\usepackage{mathtools}
\usepackage{graphicx}
\usepackage{epstopdf}
\usepackage{xfrac}
\usepackage{multirow}
\usepackage{wasysym}

\newcommand{\RA}[4]{{#1}$^\text{h}${#2}$^\text{m}${#3}$^\text{s}${#4}}
\newcommand{\dec}[4]{{#1}\textdegree {#2}'{#3}''{#4}}

\shorttitle{N/O ratio in dwarf galaxies}
\shortauthors{Douglass \& Vogeley}

\begin{document}

%
%
\title{Large-scale environmental dependence of the abundance ratio of nitrogen to oxygen in blue, star-forming galaxies fainter than $L_*$}
\author{Kelly A. Douglass, Michael S. Vogeley}
\affil{Department of Physics, Drexel University, 3141 Chestnut Street, Philadelphia, PA  19104}
\email{kelly.a.douglass@drexel.edu}

%
%
\begin{abstract}
We examine how the cosmic environment affects the chemical evolution of galaxies 
in the Universe by comparing the N/O ratio of dwarf galaxies in voids with 
dwarf galaxies in more dense regions.  Ratios of the forbidden [\ion{O}{3}] and 
[\ion{S}{2}] transitions provide estimates of a region's electron temperature 
and number density.  We estimate the abundances of oxygen and nitrogen using 
these temperature and density estimates and the emission line fluxes 
[\ion{O}{2}] $\lambda 3727$, [\ion{O}{3}] $\lambda \lambda 4959, 5007$, and 
[\ion{N}{2}] $\lambda \lambda 6548, 6584$ with the direct $T_e$ method.  Using 
spectroscopic observations from the Sloan Digital Sky Survey Data Release 7, we 
are able to estimate the N/O ratio in 42 void dwarf galaxies and 89 dwarf 
galaxies in more dense regions.  The N/O ratio for void dwarfs ($M_r > -17$) is 
slightly lower ($\sim 12\%$) than for dwarf galaxies in denser regions.   We 
also estimate the nitrogen and oxygen abundances of 2050 void galaxies and 3883 
galaxies in more dense regions with $M_r > -20$.  These somewhat brighter 
galaxies (but still fainter than $L_*$) also display similar minor shifts in the 
N/O ratio.  The shifts in the average and median element abundance values in all 
absolute magnitude bins studied are in the same direction, suggesting that the 
large-scale environment may influence the chemical evolution of galaxies.  We 
discuss possible causes of such a large-scale environmental dependence of the 
chemical evolution of galaxies, including retarded star formation and a higher 
dark matter halo mass to stellar mass ratio in void galaxies.
\end{abstract}


\maketitle

%
%
\section{Introduction}

The measurement of the abundance of heavier elements relative to hydrogen in a 
galaxy can indicate the galaxy's evolutionary stage.  As stars evolve, they 
slowly convert hydrogen into heavier elements, increasing the ratio of the 
heavier elements (oxygen, nitrogen, etc.) to hydrogen.  The ratio of oxygen to 
hydrogen is often used to determine the chemical evolution of a galaxy because 
oxygen is the most abundant element in the universe (after hydrogen and helium) 
and because oxygen has very strong emission lines in the optical regime that 
cover a range of ionization states \citep{Kewley02}.

It is instructive to also study the relative abundances of the heavy elements in 
a galaxy.  Rather than indicating the amount of hydrogen converted to heavier 
elements, the ratio of two heavy elements can reveal important details about the 
nucleosynthesis process and the chemical conditions of the galaxy when the last 
star formation episode occurred \citep{Izotov99}.  One of the easiest and most 
informative ratios to study is nitrogen to oxygen.

From what we currently understand of stellar nucleosynthesis, we can group its 
products into two classes: primary and secondary elements.  The yields of 
primary elements (carbon and oxygen, for example) are independent of the initial 
metallicity of the star, while the yields of secondary elements depend on the 
initial abundance of heavy elements in the star.  Nitrogen is unique --- it can 
behave as both a primary and secondary element \citep{Matteucci86}.  Nitrogen is 
produced during the CNO cycle, which is one of the two main processes of 
hydrogen burning in a star.  The CNO cycle fuses four protons into a helium atom 
with two positrons and two electron neutrinos as by-products.  It tends to occur 
in more massive stars than our Sun, due to the higher temperature required for 
the fusion processes involved.  Carbon is a catalyst of the CNO cycle, not a 
product.  As a result, if carbon is not initially present within the star, then 
nitrogen is produced in the same relative abundance as carbon and oxygen --- 
nitrogen behaves as a primary element.  However, if the interstellar medium 
(ISM) has a relatively high abundance of heavier elements from previous star 
formation episodes, then nitrogen behaves as a secondary element, since its 
production is based on carbon and oxygen produced prior to the star's creation.

The majority of the production of oxygen and nitrogen is thought to occur in 
different mass stars --- nitrogen is produced in the CNO cycle of 
intermediate-mass stars ($4M_{\odot} < M_* < 8M_{\odot}$), while oxygen is 
primarily produced in the helium, carbon, and neon burning stages of higher mass 
stars ($M_* > 4M_{\odot}$) \citep{Henry00,Henry06}.  The CNO cycle can occur in 
lower mass stars (the minimum temperature is only $1.5\times 10^7 K$), but it 
requires carbon as a catalyst.  If there is already carbon present in a star at 
its birth, the CNO cycle can commence much earlier in the star's lifetime than 
if it is comprised primarily of hydrogen at its birth.  

A measurement of the N/O ratio indicates where a galaxy is in its chemical 
evolution.  The relative amounts of these two elements can be influenced by 
nucleosynthesis, a galaxy's star formation history, and/or a varying initial 
mass function (IMF), for example.  The star formation history of a galaxy can be 
strongly influenced by the galaxy's environment.  Galactic interactions can 
cause bursts of star formation in addition to secular star formation.  Due to 
the time delay in the release of nitrogen and oxygen from the stellar 
population, galaxies which have more recently experienced star formation will 
result in lower N/O ratios (since oxygen is released sooner than nitrogen, due 
to higher mass stars being responsible for the production of oxygen).  In 
addition to this time delay, if a galaxy has enough heavy elements present in 
its gas at the time of the stars' births, secondary nitrogen will be produced in 
addition to primary.  This would result in higher N/O ratios, and there would be 
a correlation between the metallicity and the N/O ratio in the galaxies.

Large galaxy redshift surveys have shown that the large-scale structure of 
galaxies is similar to that of a three-dimensional cosmic web \citep{Bond96}, 
where voids (large, underdense regions that occupy approximately 60\% of space) 
separate galaxy clusters which are connected by thin filaments of galaxies.  
These cosmic voids are an important environment for studying galaxy formation 
\citep[see][for a review]{vandeWeygaert11}, as the $\Lambda$CDM cosmology 
predicts void galaxies to have lower mass and be retarded in their star 
formation when compared to those in more dense environments 
\citep[e.g.,][]{Gottlober03, Goldberg05, Cen11}.  Because dwarf galaxies are 
sensitive to many astrophysical effects including cosmological reionization, 
internal feedback from supernovae and photoheating from star formation, external 
effects from tidal interactions and ram pressure stripping, small-scale details 
of dark matter halo assembly, and properties of dark matter, they should be the 
most sensitive to the effects of the void environment.  

Previous work by \cite{Douglass17a} (hereafter referred to as Paper 1) shows 
that there is no large-scale environmental dependence of the amount of oxygen in 
dwarf galaxies, in contrast to earlier studies by \cite{Pustilnik06, Cooper08, 
Deng11, Filho15}, for example.  One of the main arguments for the existence of 
an environmental dependence of the metallicity of galaxies centers around the 
idea that void galaxies are surrounded by pristine hydrogen that is unavailable 
to galaxies in more dense regions.  By looking at just N/O, we remove the 
hydrogen dependence of the relative abundances.  Detecting a difference in the 
N/O ratio due to the large-scale environment would indicate that the cosmic 
environment has some influence on the nucleosynthesis of secondary elements.  In 
addition, if the environment does have some very minor effect on the metallicity 
of a galaxy, removing the hydrogen dependence could amplify this effect above 
the noise of the data.  Combined with the metallicity results in Paper 1, we 
might be able to discern a large-scale environmental effect on the chemical 
evolution of galaxies.

Large-scale sky surveys like the Sloan Digital Sky Survey \citep{Abazajian09} 
contain a large sample of dwarf galaxies, allowing us to analyze the dwarf 
galaxy population in the nearby universe with more statistical significance.  
Over 1000 voids have been identified in SDSS DR7 \citep{Pan12}, and SDSS 
provides spectroscopy to permit abundance estimates of those dwarf galaxies 
found in these voids.  Thus, we are able to estimate the N/O ratio as a function 
of large-scale environment for the largest sample of dwarf galaxies to date.

We make use of the MPA-JHU catalog's reprocessed spectroscopic 
data\footnote{Available at http://www.mpa-garching.mpg.de/SDSS/DR7/} to study 
the N/O abundance ratio of a large collection of dwarf galaxies in SDSS DR7.  
Because our analysis depends on the weak [\ion{O}{3}] $\lambda 4363$ auroral 
line, the MPA-JHU catalog's more detailed treatment of the stellar continuum 
permits the weaker emission lines to become more apparent.  As a result, using 
this catalog's flux measurements should improve the accuracy of our results.  We 
study the N/O abundance ratio of these dwarf galaxies as a function of 
large-scale environment to discern if the large-scale environment has an effect 
on the relative abundance of heavier elements in dwarf galaxies.

Our paper is organized as follows.  Section 2 describes the method used to 
estimate the chemical abundances in galaxies.  We remind the reader of the 
source of our data in Section 3.  Section 4 includes the results of our analysis, 
and Section 5 is a discussion of the implications of our results on the 
large-scale environmental effects on galaxy evolution.  Finally, Section 6 
summarizes our conclusions and discusses future work.

%
%
\section{Estimation of gas-phase chemical abundances from optical spectroscopy}
\label{sec:Theory}

We study a galaxy's oxygen and nitrogen abundances because they are relatively 
abundant elements, they emit strong lines in the optical regime (including for 
several ionization states in oxygen), and a ratio of some of the oxygen lines 
provides a good estimate of the electron temperature \citep{Kewley02}.  What 
follows is a description of the theory and methods we employ to estimate the 
oxygen and nitrogen abundances in dwarf galaxies.

\subsection{[\ion{N}{2}]}

\begin{figure}
    \centering
    \includegraphics[width=0.47\textwidth]{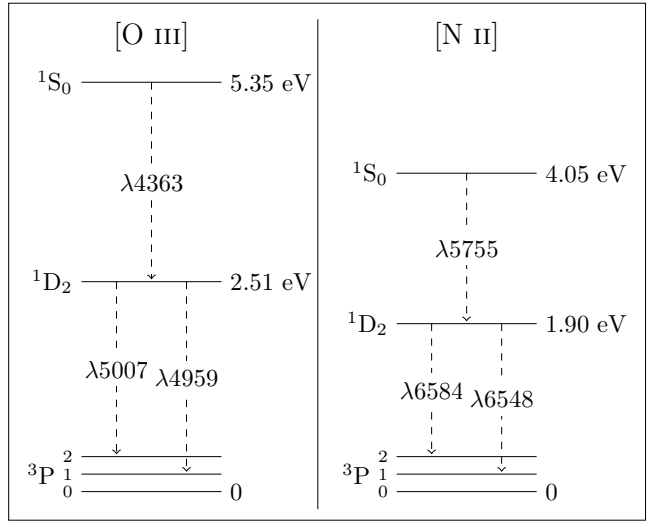}
    \caption{Energy-level diagram for [\ion{O}{3}] and [\ion{N}{2}] $(2p^2)$ ions.  
    The most important transitions are shown; all are in the visible spectrum.  
    These forbidden transitions in both oxygen and nitrogen provide an estimate 
    of the electron temperature in the interstellar gas.  Because oxygen is more 
    abundant, we use the oxygen lines to estimate the temperature of the gas.}
    \label{fig:transitions}
\end{figure}

The energy level diagram for the various transitions of [\ion{N}{2}] is very 
similar to that of [\ion{O}{3}], since they have the same electron ground state 
configuration ((1s)$^2$(2s)$^2$(2p)$^2$).  The similarities can be seen in Fig. 
\ref{fig:transitions}.  Therefore, an estimate of the electron temperature can 
be made from the [\ion{N}{2}] $\lambda 5755$ emission line.  However, this line 
is weaker than the [\ion{O}{3}] $\lambda 4363$ auroral line (since there is less 
N than O in galaxies), so we use the [\ion{O}{3}] auroral line for our 
temperature estimates, as in Paper 1.  After obtaining a temperature and density 
estimate, we use the [\ion{N}{2}] $\lambda \lambda 6548, 6584$ doublet to 
estimate the abundance of singly ionized nitrogen in a galaxy.

\subsection{Direct $T_e$ method}

We use the same method to calculate the nitrogen abundance as in Paper 1 to 
estimate the oxygen abundance.  However, here we use the [\ion{N}{2}] 
$\lambda \lambda 6548, 6584$ doublet instead of the [\ion{O}{2}] $\lambda 3727$ 
and [\ion{O}{3}] $\lambda \lambda 4959,5007$ doublets.  Because the temperature 
estimate depends on the auroral line [\ion{O}{3}] $\lambda 4363$, this method is 
often difficult to employ.  As a result, it works best with low redshift, 
low metallicity galaxies.  The electron temperature is derived by solving the 
following system of equations:
\begin{equation}
	t_3 = \frac{1.432}{\log[(\lambda 4959 + \lambda 5007)/\lambda 4363] - \log C_T}
\end{equation}
where $t_3 = 10^{-4} T_e(\text{O}^{++})$ and
\begin{equation}
	C_T = (8.44 - 1.09t_3 + 0.5t_3^2 - 0.08t_3^3)\frac{1 + 0.0004x_3}{1 + 0.044x_3}
\end{equation}
where $x_3 = 10^{-4} n_e t_3^{-0.5}$.  The ionic abundances are then found with 
the equations
\begin{align}
    12 + \log \left( \frac{\text{O}^{++}}{\text{H}^+} \right) &= \log \frac{\lambda 4959 + \lambda 5007}{\text{H}\beta} + 6.200 \nonumber \\
	&\qquad + \frac{1.251}{t_3} - 0.55\log t_3 \nonumber \\
	&\qquad - 0.014t_3 \\
	12 + \log \left( \frac{\text{O}^+}{\text{H}^+} \right) &= \log \frac{\lambda 3727}{\text{H}\beta} + 5.961 + \frac{1.676}{t_2} \nonumber \\
	&\qquad - 0.40\log t_2 - 0.034t_2 \nonumber \\
	&\qquad + \log (1+1.35x_2) \\
	12 + \log \left( \frac{\text{N}^+}{\text{H}^+} \right) &= \log \frac{\lambda 6548 + \lambda 6584}{\text{H}\beta} + 6.234 \nonumber \\
	&\qquad + \frac{0.950}{t_2} - 0.42\log t_2 - 0.027t_2 \nonumber \\
	&\qquad + \log (1 + 0.116x_2)
\end{align}
where $t_2 = 10^{-4} T_e(\text{O}^+)$ and $x_2 = 10^{-4} n_e t_2^{-0.5}$.  We 
assume that $T_e(\text{N}^+) = T_e(\text{O}^+)$.

The S/N of the SDSS spectra is too low to directly estimate the temperature of 
the gas in the low-ionization zone.  As a result, we use the relation 
$t_2 = 0.7t_3 + 0.3$ by \cite{Garnett92}.  This relation has been shown to 
overestimate this temperature \citep{Andrews13}.  Since the metal emission lines 
are the primary method of cooling for the gas, a high temperature corresponds to 
a low metallicity.  Therefore, an overestimate of the temperature results in an 
underestimated abundance.  As shown in Paper 1, this only affects perhaps 15 of 
the dwarf galaxies in our sample and does not influence our conclusions.

The sum of the abundances of each of the element's ionization states is equal to 
the total abundance of any element, whether or not all ionization states are 
observed.  Most of oxygen exists as either singly or doubly ionized, so the 
total oxygen abundance is
\begin{equation}
	\frac{\text{O}}{\text{H}} = \frac{\text{O}^{++}}{\text{H}^+} + \frac{\text{O}^+}{\text{H}^+}
\end{equation}
Since we can only observe the nitrogen abundance in one of the main ionization 
states, we use an ionization correction factor (ICF) to account for the missing 
states.  For any element $X$, the total abundance is 
\begin{equation}\label{eqn:ICF}
    \frac{\text{X}}{\text{H}} = \sum_i ICF_i \frac{\text{X}^i}{\text{H}}
\end{equation}
For nitrogen, we employ the ICFs as defined in \cite{Izotov06}:
\begin{equation}\label{eqn:ICF_N}
    ICF(\text{N}^+) = \left\{ \begin{array}{ll}
    -0.825v + 0.718 + \frac{0.853}{v} & \mbox{low $Z$}\\
    -0.809v + 0.712 + \frac{0.852}{v} & \mbox{intermed $Z$}\\
    1.467v + 1.752 + \frac{0.688}{v} & \mbox{high $Z$}
    \end{array} \right.
\end{equation}
where $v = \text{O}^+ / (\text{O}^+ + \text{O}^{++})$.  The range for low $Z$ 
covers galaxies with $12 + \log \left( \text{O}/\text{H} \right) \leq 7.2$, 
while high $Z$ includes galaxies with 
$12 + \log \left( \text{O}/\text{H} \right) \geq 8.2$.  For galaxies with 
$7.2 < 12 + \log \left( \text{O}/\text{H} \right) < 7.6$, the values for the 
$ICF$s are a linear interpolation between the low $Z$ and intermediate $Z$ 
values, while the $ICF$s for galaxies with 
$7.6 < 12 + \log \left( \text{O}/\text{H} \right) < 8.2$ are a linear 
interpolation between the intermediate $Z$ and high $Z$ values.

The N/O ratio can be found from the O/H and N/H ratios:
\begin{equation}\label{eqn:NO}
    \log \left(\frac{\text{N}}{\text{O}}\right) = \left[ 12 + \log \left( \frac{\text{N}}{\text{H}} \right) \right] - \left[ 12 + \log \left( \frac{\text{O}}{\text{H}} \right) \right]
\end{equation}

%
%
\section{SDSS data and galaxy selection}

The SDSS Data Release 7 \citep[DR7;][]{Abazajian09} uses drift scanning to map 
approximately one-quarter of the northern sky; it is a wide-field multi-band 
imaging and spectroscopic survey.  A dedicated 2.5 m telescope at the Apache 
Point Observatory in New Mexico \citep{Fukugita96, Gunn98} takes the photometric 
data in the five band SDSS system --- $u$, $g$, $r$, $i$, and $z$.  Galaxies 
selected for spectroscopic analysis must have a Petrosian $r$-band magnitude 
$m_r < 17.77$ \citep{Lupton01, Strauss02}.  Two double fiber-fed spectrographs 
and fiber plug plates take the spectra in an observed wavelength range of 
3800--9200 \AA with a resolution $\lambda / \Delta \lambda \sim 1800$ and a 
minimum fiber separation of 55 arcsec \citep{Blanton03}.  As in Paper 1, we use 
the emission line flux data from the MPA-JHU value-added catalog, which is based 
on the SDSS DR7 sample of galaxies.  Total star formation rates and total 
specific star formation rates are also from the MPA-JHU value-added catalog, 
following the technique discussed in \cite{Brinchmann04}.  The MPA-JHU catalog 
is also the source of the stellar mass estimates used, as calculated in 
\cite{Tremonti04}, following the method outlined in \cite{Kauffmann03}.  The 
KIAS value-added galaxy catalog \citep{Choi10} is our source of the absolute 
magnitudes and colors of the galaxies.

\subsection{Spectroscopic selection}\label{sec:SDSS_limits}

The following requirements are implemented on the SDSS DR7 main spectroscopic 
galaxy sample described above.  We use the same requirements for our sample as 
in Paper 1: all galaxies must have
\begin{itemize}
    \item{$M_r > -17$ (dwarf galaxies)}
    \item{a minimum $5\sigma$ detection of H$\beta$}
    \item{a minimum $1\sigma$ detection of [\ion{O}{3}] $\lambda 4363$}
    \item{a flux $>0$ for all other required lines}
    \item{$T_e (\text{\ion{O}{3}}) < 3\times 10^4 \text{ K}$}
    \item{a star-forming BPT classification by \cite{Brinchmann04}}
\end{itemize}
We also use the \texttt{oii\_flux} value from the MPA-JHU catalog in place of 
their [\ion{O}{2}] $\lambda \lambda 3726, 3729$ flux measurement since we are 
working at such low redshifts ($0.02 < z < 0.03$).  Detailed descriptions of 
these criteria can be found in Section 3.1 of Paper 1.

\subsection{Void classification}

The large-scale environment of the galaxies was determined using the void 
catalog constructed by \cite{Pan12}, which is based on the galaxies in the SDSS 
DR7 catalog.  The VoidFinder algorithm of \cite{Hoyle02} removes all isolated 
galaxies with absolute magnitudes $M_r < -20$ (a galaxy is defined to be 
isolated if its third nearest neighbor is more than 7 $h^{-1}$ Mpc away).  
Placing a grid over the remaining galaxies, VoidFinder grows spheres in the 
centers of all grid cells that contain no galaxies.  The spheres expand until 
they encounter four galaxies on the surface.  To be considered part of a void, a 
sphere must have a minimum radius of 10 Mpc; two spheres which overlap by more 
than 10\% are considered part of the same void.  We refer the reader to 
\cite{Hoyle02} for a more detailed description of the VoidFinder algorithm.  
Using these voids, galaxies that live within any void sphere are classified as 
a void galaxy; those which are outside the spheres are considered wall galaxies.  
Due to the construction of the void spheres, we cannot identify any voids within 
10 Mpc of the edge of the survey.  As a result, the large-scale environment of 
any galaxy within this boundary is uncertain.

9519 of the $\sim$800,000 galaxies with spectra available in SDSS DR7 are dwarf 
galaxies ($M_r > -17$).  42 void dwarf galaxies, 89 wall dwarf galaxies, and 4 
dwarf galaxies with uncertain large-scale environments are left to analyze after 
applying the spectroscopic cuts (or 135 dwarf galaxies in total, 131 of which 
are used in the environmental study).

%
%
\section{Abundance analysis and results}

Our primary objective is to perform a relative measurement of the N/O ratio of 
dwarf galaxies to discern how the large-scale environment affects their chemical 
evolution.  As discussed in Paper 1, multiple methods have been developed for 
metallicity calculations based on the quality of the spectra.  We use only the 
direct $T_e$ method for our abundance calculations due to the limited galaxy 
types used in the calibration or theoretical development of other methods.

For reference, the solar metallicity $Z_{\odot} = 8.69\pm 0.05$ 
\citep{Asplund09}.

\subsection{Estimation of uncertainties and comparison of N/O and N$^+$/O$^+$}

We estimate uncertainties in the computed abundances using a Monte Carlo method.  
We calculate 100,000 abundance estimates using the measured line fluxes and 
scaled uncertainty estimates.  A new positive ``fake'' line flux is drawn from a 
normal distribution for each abundance estimate.  The standard deviation in 
the sets of 100,000 calculated abundance values is used for the error in the 
abundance calculation.  A more in-depth description of this process can be found 
in Paper 1.

\begin{figure}
    \centering
    \includegraphics[width=0.5\textwidth]{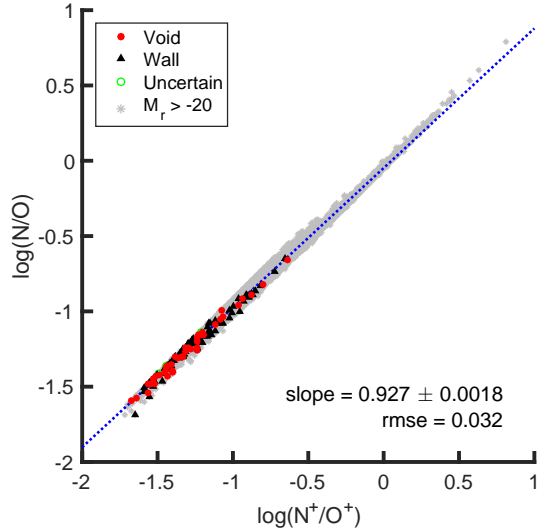}
    \caption{A comparison of the N/O and N$^+$/O$^+$ abundance ratios for our 
    dwarf galaxies.  Brighter galaxies ($M_r > -20$) are shown in gray for 
    context.  All our star-forming galaxies with $M_r > -20$ roughly follow the 
    approximation that N/O $\cong$ N$^+$/O$^+$, which is often assumed in other 
    studies of the abundance ratio of nitrogen to oxygen.  In this paper, we 
    will be using the abundance ratio N/O for our analysis.}
    \label{fig:NO_NpOp}
\end{figure}

It has been common practice to assume that N/O $\cong$ N$^+$/O$^+$, thus 
eliminating the need for the ICF in Eqn. \ref{eqn:ICF}.  We find that this is a 
reasonable but slightly biased approximation, agreeing with the results of 
\cite{Nava06}.  A comparison of the N/O ratio and the N$^+$/O$^+$ ratio for our 
set of dwarf galaxies can be seen in Fig. \ref{fig:NO_NpOp}; galaxies with 
absolute magnitudes $M_r > -20$ are shown in grey for context.  A linear fit to 
all star-forming galaxies with magnitudes $M_r > -20$ has a slope of only 
$0.927\pm 0.0018$, with a root mean square error of 0.032 for the fit.  This 
comparison indicates that lower values of the N$^+$/O$^+$ ratio underestimate 
the N/O ratio, while higher values of N$^+$/O$^+$ overestimate the N/O ratio.  
Throughout this paper, we study the N/O ratio using ICF-corrected estimates of 
N/O (Eqns. \ref{eqn:ICF_N} and \ref{eqn:NO}).

\subsection{Sources of systematic error}

There is a radial dependence of many physical properties of galaxies 
\citep{Bell00}.  Consequently, abundance estimates may depend on the locations 
of the spectroscopic fiber on the galaxy.  If all of the galaxy's light is not 
contained within the fiber of the spectrograph, the estimated abundances will 
not necessarily be representative of global abundance values.  For example, 
\cite{Bell00} show that the metallicity is not constant throughout a galaxy.  
Due to the spatially resolved spectra produced by MaNGA of SDSS-IV 
\citep{SDSS13}, a statistically significant measure of the radial dependence of 
a galaxy's metallicity should soon be possible \citep{Wilkinson15}.  In SDSS 
DR7, the fiber diameter is 3 arcseconds, corresponding to a physical diameter 
between 1.29 kpc and 1.93 kpc at redshifts $0.02 < z < 0.03$.  This covers a 
majority of most dwarf galaxies' luminous surfaces.  The fiber is almost always 
placed on the brightest spot of the galaxy, which is often the center of the 
galaxy for spiral and ellipticals.  Since the metallicity has been shown to 
decrease at large radius, these abundance values may be overestimates of the 
global abundances.  Since many dwarf galaxies are irregular galaxies, the fiber 
is instead focused on a bright \ion{H}{2} region.  As a result, we are 
estimating the abundances of the gas from which stars recently formed.

We are implicity limiting our sample of galaxies to only blue, star-forming 
dwarf galaxies as a result of our selection criteria outlined in Section 
\ref{sec:SDSS_limits}.  Consequently, this is not a representative sample of the 
full dwarf galaxy population.  In this study we are only able to discuss the 
large-scale environmental influence on blue, star-forming dwarf galaxies within 
a narrow redshift range.  It is impossible to use the direct $T_e$ method to 
measure the chemical abundances of red dwarf galaxies because the UV photons 
from young stars are needed to excite the interstellar gas.

\subsection{Galaxy abundances}

Abundances estimated using the direct $T_e$ method for our dwarf galaxy sample 
are listed in Table \ref{tab:Results}, along with other important 
characteristics and identification for the galaxies (including their large-scale 
environment classification).

\floattable
\input{t1_short}

\subsubsection{Oxygen, nitrogen abundances}\label{sec:OH_NH}

\begin{figure*}
    \centering
    \includegraphics[width=0.49\textwidth]{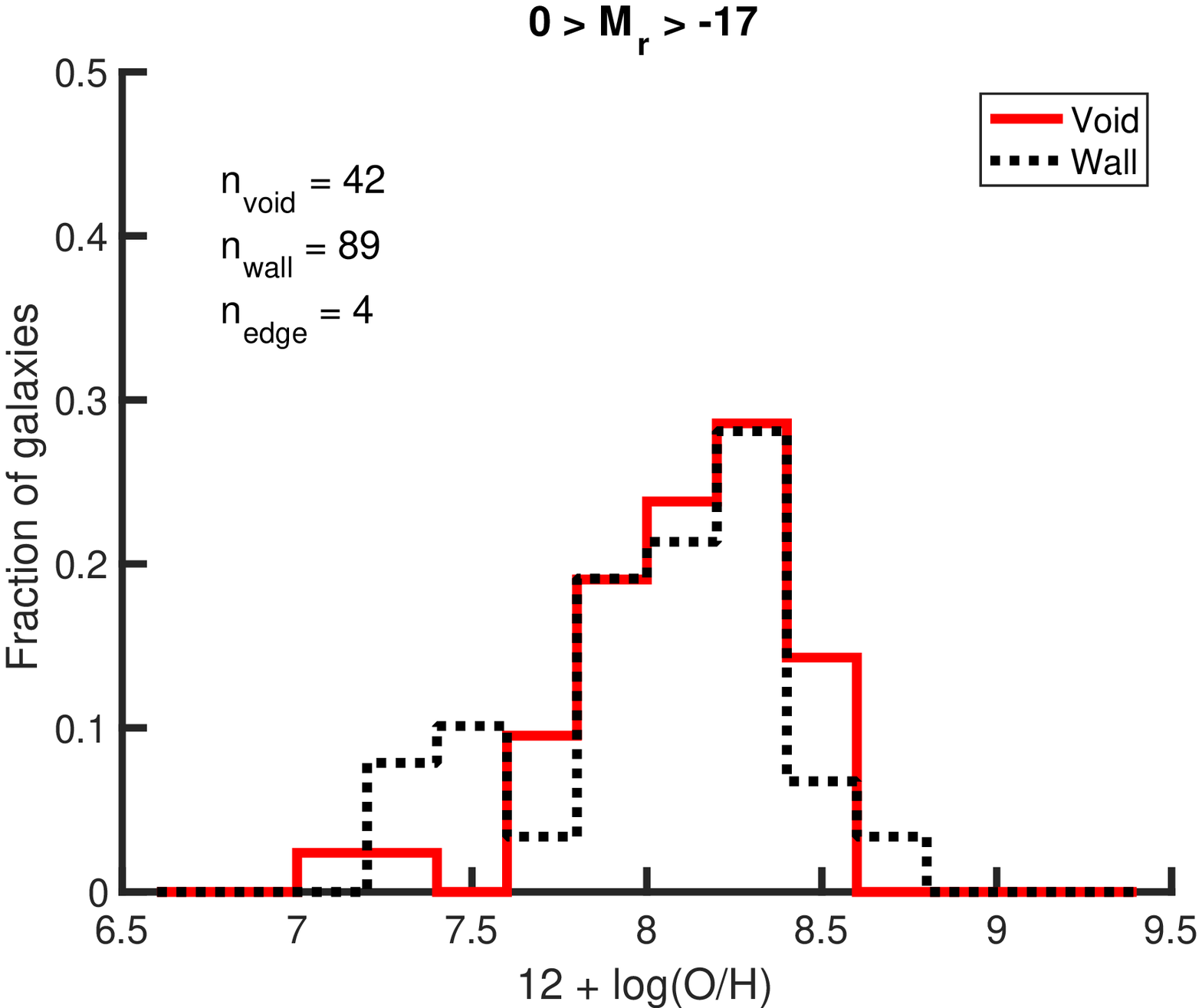}
    \includegraphics[width=0.49\textwidth]{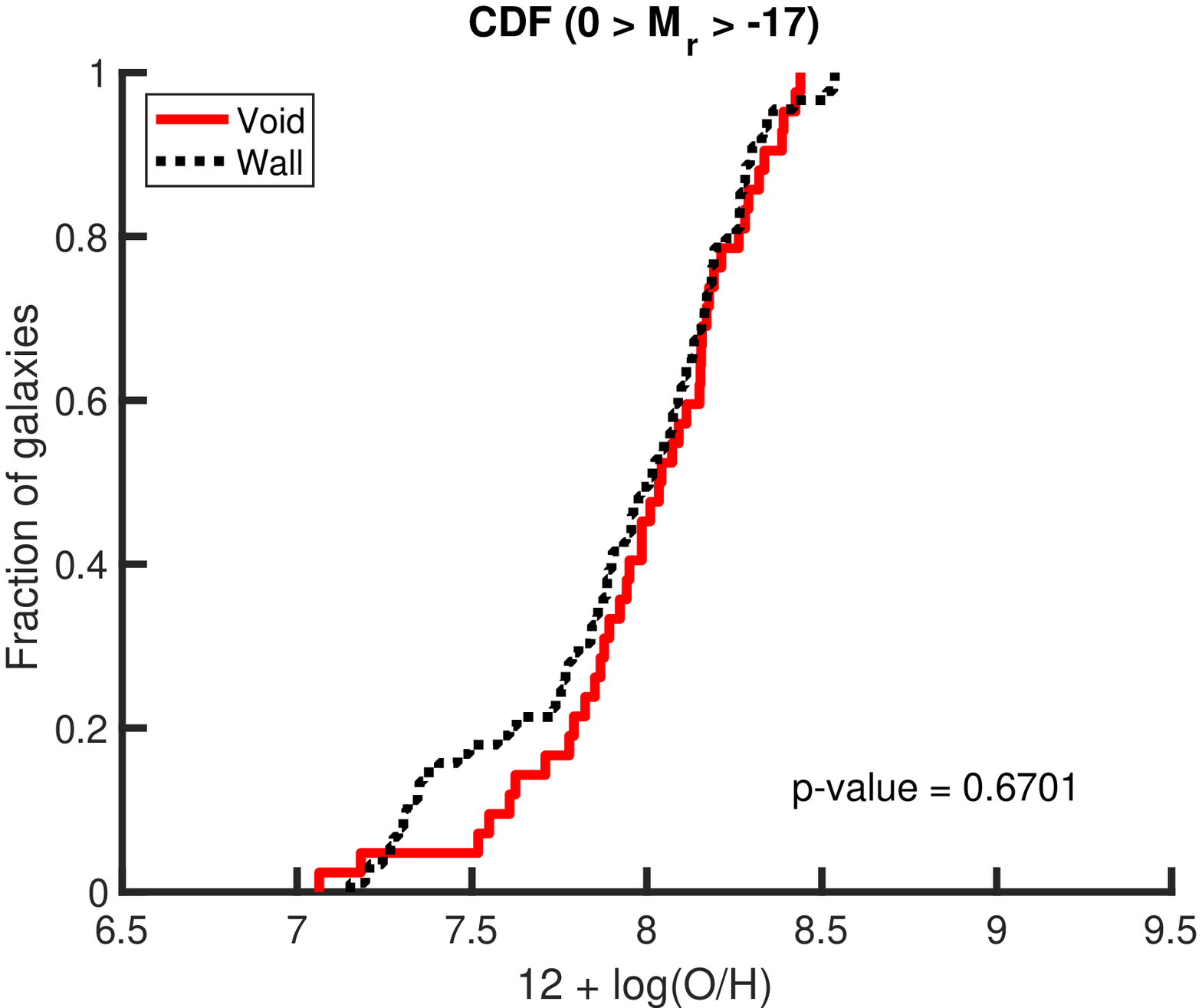}
    \caption{Gas-phase oxygen abundances (relative to hydrogen) of void dwarf 
    (red solid line) and wall dwarf (black dashed line) galaxies, taken from 
    Fig. 4 in Paper 1.  As we demonstrate in Paper 1, the gas-phase oxygen 
    abundances in dwarf galaxies do not depend on the large-scale environment.}
    \label{fig:met1sig}
\end{figure*}

\begin{figure*}
    \centering
    \includegraphics[width=0.49\textwidth]{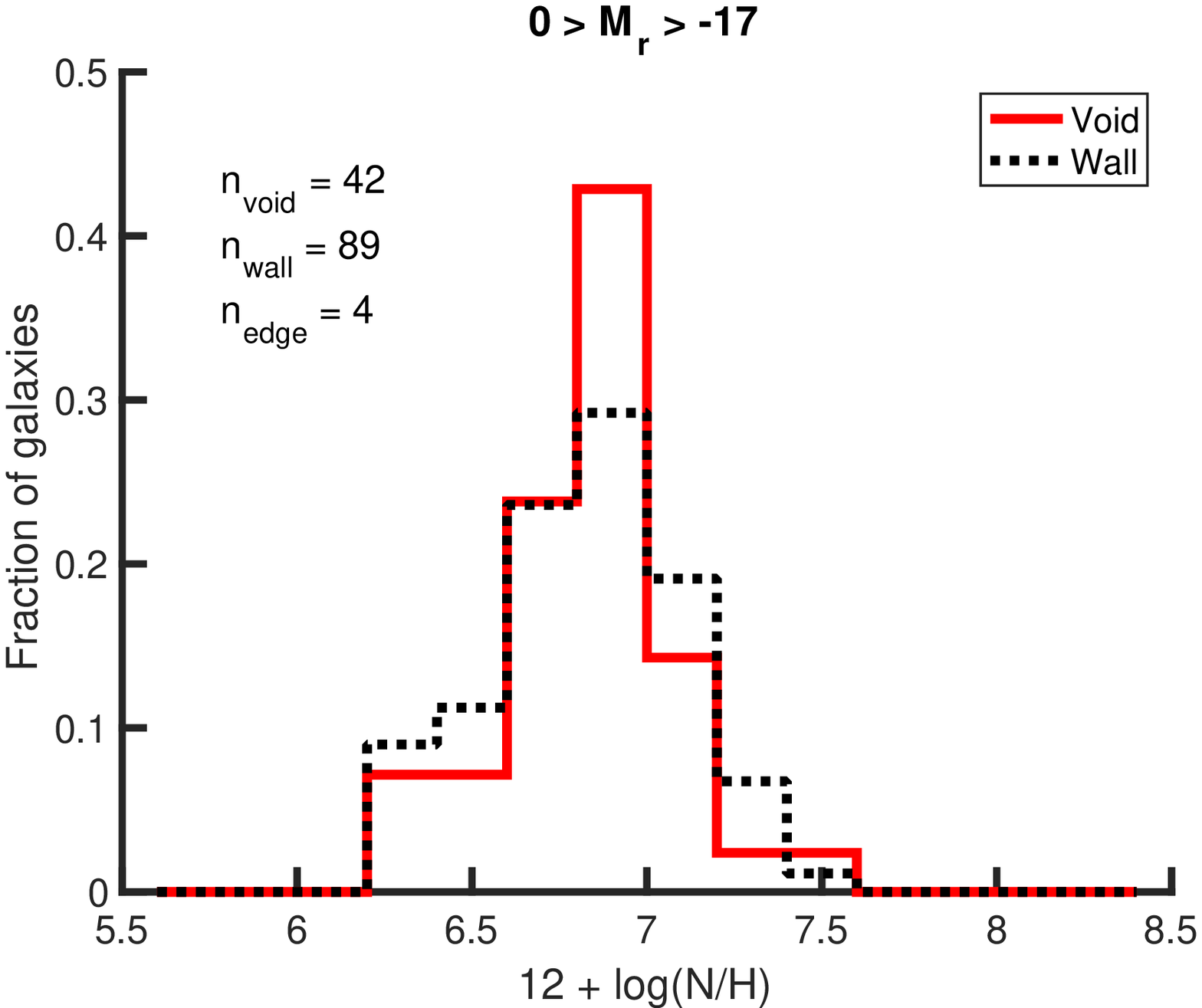}
    \includegraphics[width=0.49\textwidth]{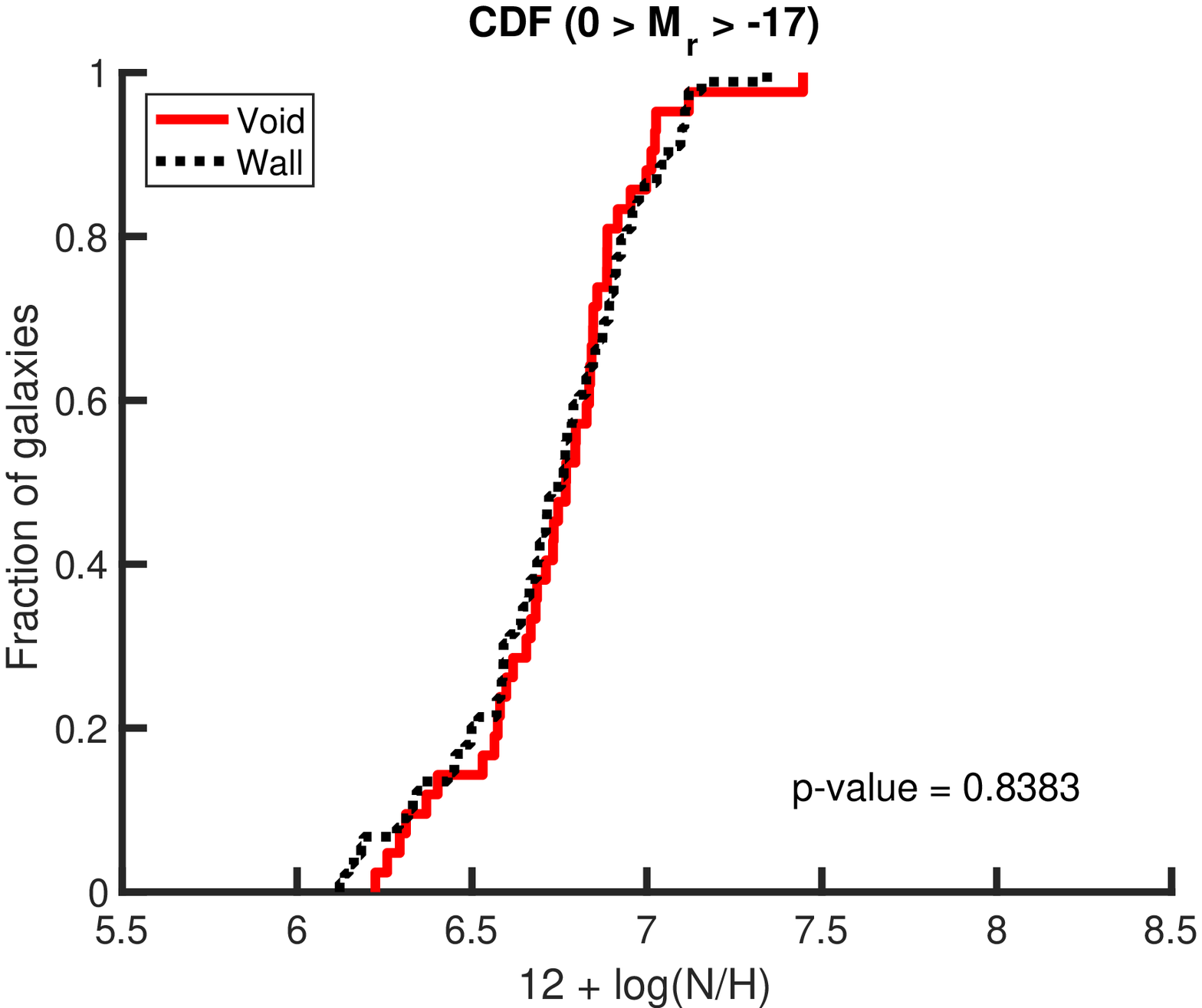}
    \caption{The abundance of nitrogen relative to hydrogen of void dwarf (red 
    solid line) and wall dwarf (black dashed line) galaxies.  A two-sample KS 
    test of the two data sets results in an asymptotic $p$-value of 0.84, 
    indicating an 84\% probability that a test statistic greater than the 
    observed value of 0.11 will be seen if the void sample is drawn from the 
    wall sample.  This is reflected visually, as there appears to be very little 
    difference between the two populations, indicating that there is little 
    large-scale environmental influence on the nitrogen abundance of dwarf 
    galaxies.}
    \label{fig:N_1sig}
\end{figure*}

Histograms of the resulting oxygen and nitrogen abundances are shown in Figs. 
\ref{fig:met1sig} and \ref{fig:N_1sig}, respectively.  Both figures show very 
little difference in the distribution of abundance values in dwarf galaxies 
between voids and walls.  A two-sample Kolmogorov-Smimov (KS) test quantifies 
this observation --- it produced a test statistic of 0.13 for oxygen and 0.11 
for nitrogen, corresponding to a probability of 67.1\% and 83.8\%, respectively, 
that a test statistic greater than or equal to this calculated test statistic 
will be measured if the void sample were drawn from the wall sample.  The 
cumulative distribution function (CDF) of these samples can be seen on the right 
in Figs. \ref{fig:met1sig} and \ref{fig:N_1sig}.  The KS test quantifies the 
visual impression in these figures that the distributions of oxygen and nitrogen 
abundances are similar for dwarf galaxies in voids and walls.

The average and median values of the dwarf galaxy abundances indicate very 
little large-scale environmental influence on the oxygen and nitrogen 
abundances.  The average oxygen abundance for void dwarf galaxies is 
$7.99\pm 0.049$ and the median is 8.04, while the average for wall dwarf 
galaxies is $7.93\pm 0.036$ with a median value of 8.01.  This implies that the 
wall dwarf galaxies have lower oxygen abundances by an average of 
$0.07\pm 0.060$ relative to the void dwarf galaxies; the shift in the median 
values is 0.03 for the dwarf galaxies, with wall dwarf galaxies having lower 
oxygen abundances than void dwarf galaxies.  There is a also a shift in the 
nitrogen abundances for the dwarf galaxies: void dwarf galaxies have an average 
nitrogen abundance of $6.74\pm 0.035$ and a median of 6.77, while the wall dwarf 
galaxies have an average nitrogen abundance of $6.72\pm 0.025$ and a median of 
6.75.  Again, wall dwarf galaxies have, on average, $0.02\pm 0.043$ lower 
nitrogen abundances than the void dwarf galaxies (the median shift for the 
nitrogen abundance of dwarf galaxies is 0.01, with wall galaxies lower than void 
dwarf galaxies).  These shifts are within the uncertainty, so they are not 
statistically significant --- if there is a large-scale environmental influence 
on the abundances of oxygen and nitrogen relative to hydrogen in dwarf galaxies, 
it is small.

\begin{figure}
    \centering
    \includegraphics[width=0.5\textwidth]{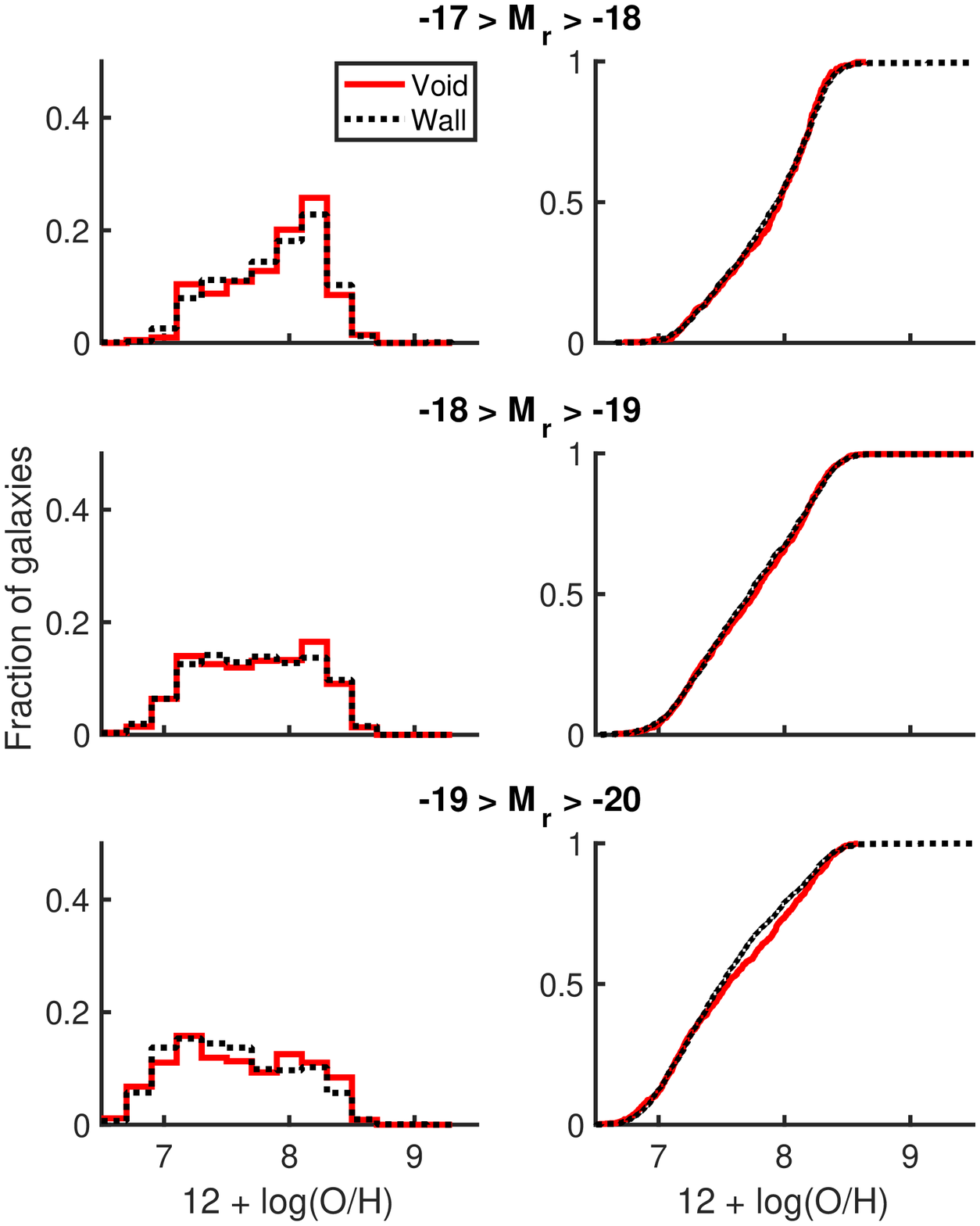}
    \caption{Gas-phase oxygen abundances relative to hydrogen of void (red 
    solid line) and wall (black dashed line) star-forming galaxies with 
    $-17 > M_r > -18$ (top), $-18 > M_r > -19$ (middle), and $-19 > M_r > -20$ 
    (bottom).  The results of a two-sample KS test of the two data sets in each 
    absolute magnitude range can be found in Table \ref{tab:stats}.  These 
    results are reflected visually, as there appears to be very little 
    difference between the two populations (regardless of absolute magnitude), 
    indicating that there is little large-scale environmental influence on the 
    oxygen abundance of star-forming galaxies with absolute magnitudes 
    $-17 > M_r > -20$.}
    \label{fig:Z_bright}
\end{figure}

\begin{figure}
    \centering
    \includegraphics[width=0.5\textwidth]{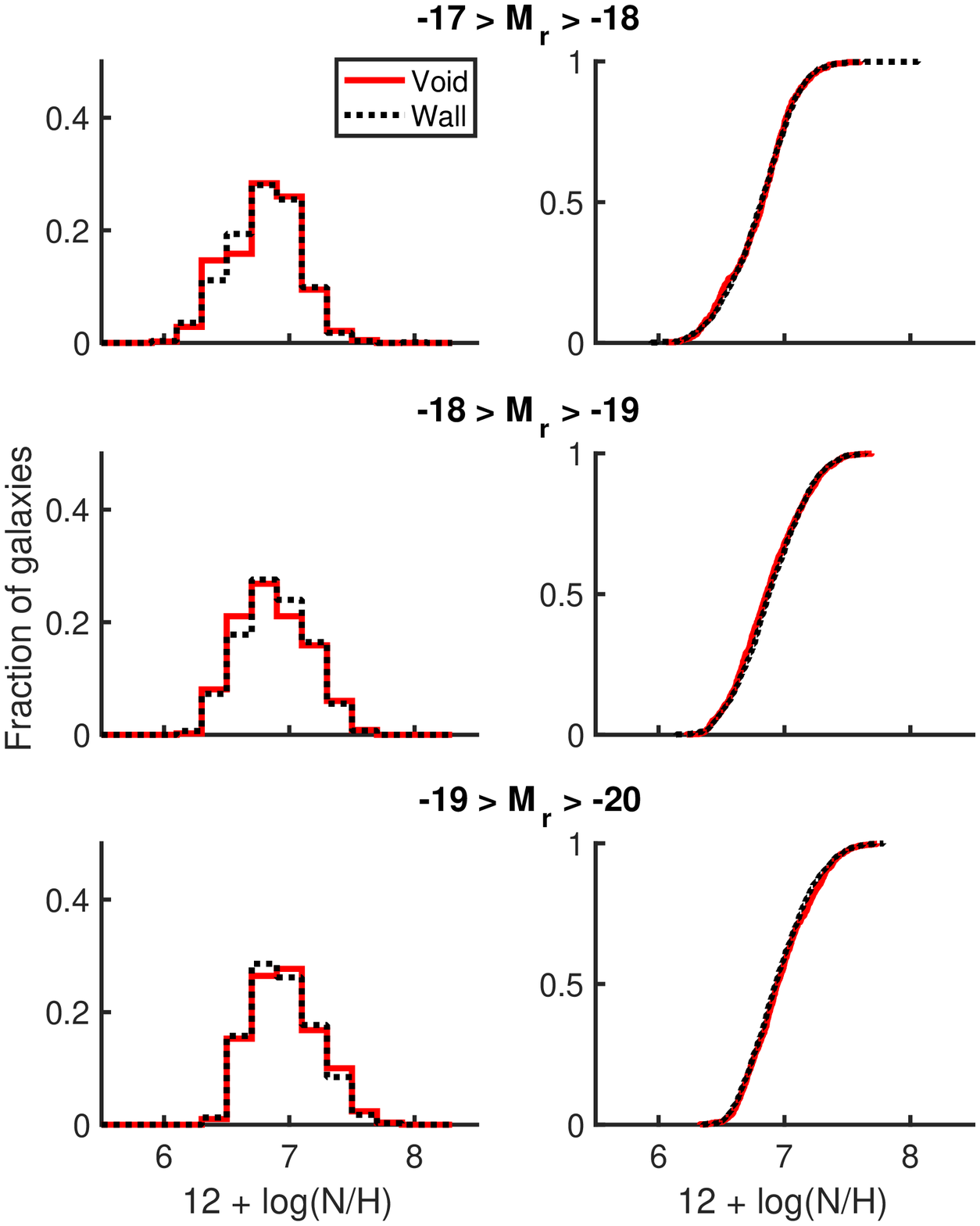}
    \caption{The abundance of nitrogen relative to hydrogen of void (red solid 
    line) and wall (black dashed line) star-forming galaxies with 
    $-17 > M_r > -18$ (top), $-18 > M_r > -19$ (middle), and $-19 > M_r > -20$ 
    (bottom).  The results of a two-sample KS test of the two data sets in each 
    absolute magnitude range can be found in Table \ref{tab:stats}.  These 
    results are reflected visually, as there appears to be very little 
    difference between the two populations (regardless of absolute magnitude), 
    indicating that there is little large-scale environmental influence on the 
    nitrogen abundance of star-forming galaxies with absolute magnitudes 
    $-17 > M_r > -20$.}
    \label{fig:N_bright}
\end{figure}

To see how our results of the environmental dependence of dwarf galaxies compare 
with somewhat brighter galaxies, we perform the same analysis on galaxies with 
absolute magnitudes $-17 > M_r > -20$.  The results of this analysis can be seen 
in Figs. \ref{fig:Z_bright} and \ref{fig:N_bright}.  As the dwarf galaxies have 
already shown, there is no obvious large-scale environmental dependence of the 
oxygen and nitrogen abundances of these brighter galaxies.  The results of a 
two-sample KS test (listed in Table \ref{tab:stats}) mostly support this 
conclusion.  In the brightest magnitude bin (galaxies with $-19 > M_r > -20$), 
the KS test returns a p-value of only 0.00062 for the oxygen abundances, 
indicating only a 0.062\% chance that there will be a test statistic greater 
than 0.07 if the void sample is drawn from the wall sample.  The oxygen 
abundances for void galaxies are higher than the wall galaxies by an average of 
$0.04\pm 0.017$ in this absolute magnitude bin, reinforcing the results of the 
KS test that there is a large-scale environmental influence on the oxygen 
abundance in galaxies with magnitudes $-19 > M_r > -20$.  While the results of 
the KS test are not as convincing for this magnitude range in the nitrogen 
abundances, there is still an average shift of $0.02\pm 0.011$ towards higher 
nitrogen abundances for void galaxies.  While only one magnitude bin shows a 
statistically significant shift between the two environments, all magnitude bins 
are shifted in the same direction.  This trend suggests that there may be a mild 
influence on the chemical evolution of galaxies due to their large-scale 
environment.

We note that there appears to be a shift in the oxygen and nitrogen abundances 
between the absolute magnitude bins in Figs. \ref{fig:Z_bright} and 
\ref{fig:N_bright} that is opposite to what is predicted by the mass-metallicity 
relation \citep{Tremonti04}.  This shift towards lower oxygen abundances as the 
galaxies increase in brightness is possibly due to the fact that the metallicity 
estimates are so dependent on the temperature-sensitive [\ion{O}{3}] 
$\lambda 4363$ auroral line.  As galaxies increase in metallicity, this line 
becomes weaker (as its strength is inversely proportional to the temperature).  
If the flux of this line is being underestimated then the temperature is being 
overestimated, and therefore the oxygen and nitrogen abundances are being 
underestimated.  If we are seeing an underestimate of flux of the [\ion{O}{3}] 
$\lambda 4363$ emission line (and therefore an overestimate of the temperature 
in the region), then we should see a shift towards lower N/H values as the 
absolute magnitude is increased as well.  This pattern can be seen in Fig. 
\ref{fig:N_bright}.

\subsubsection{Ratio of nitrogen to oxygen}

\begin{figure*}
    \centering
    \includegraphics[width=0.49\textwidth]{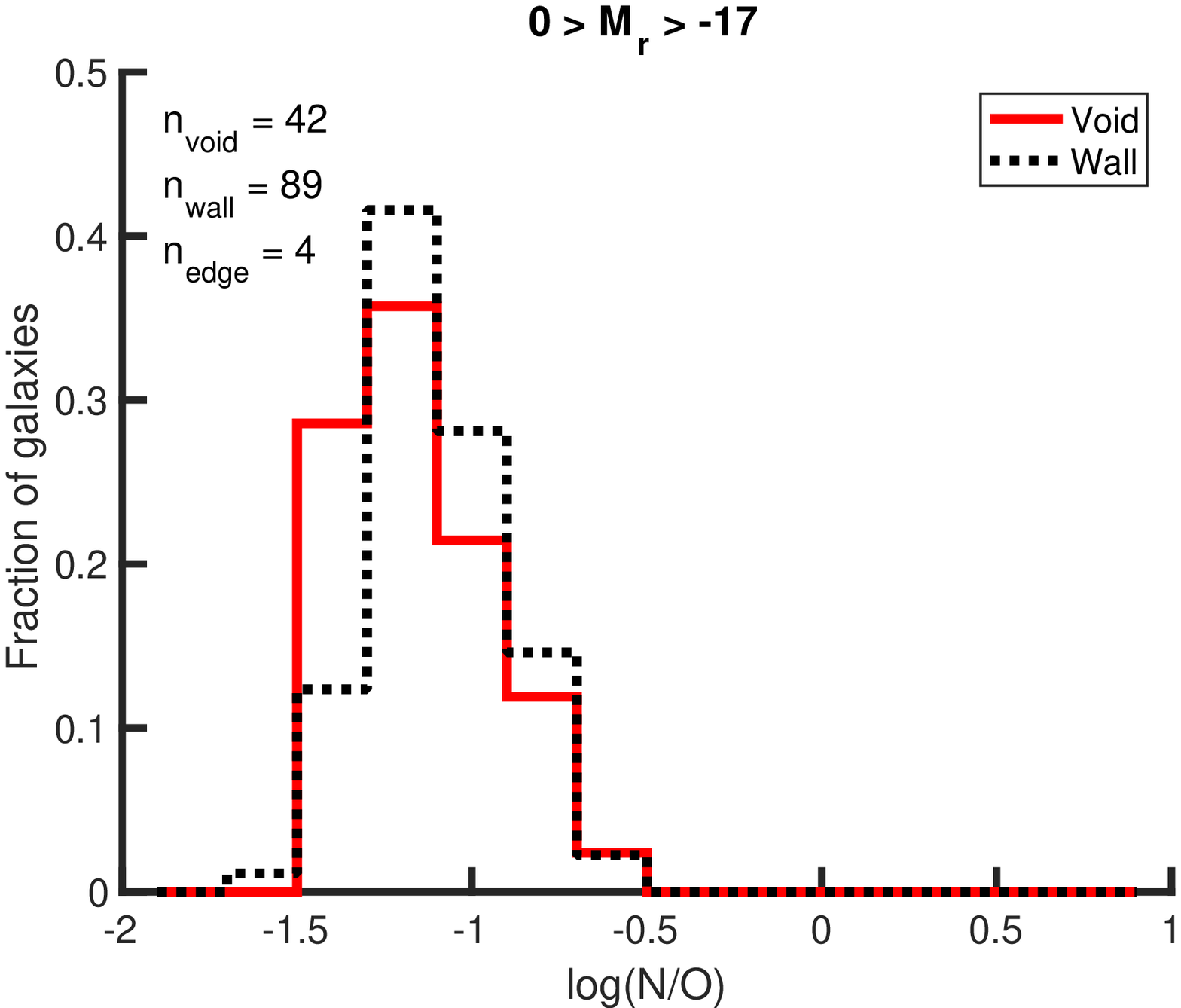}
    \includegraphics[width=0.49\textwidth]{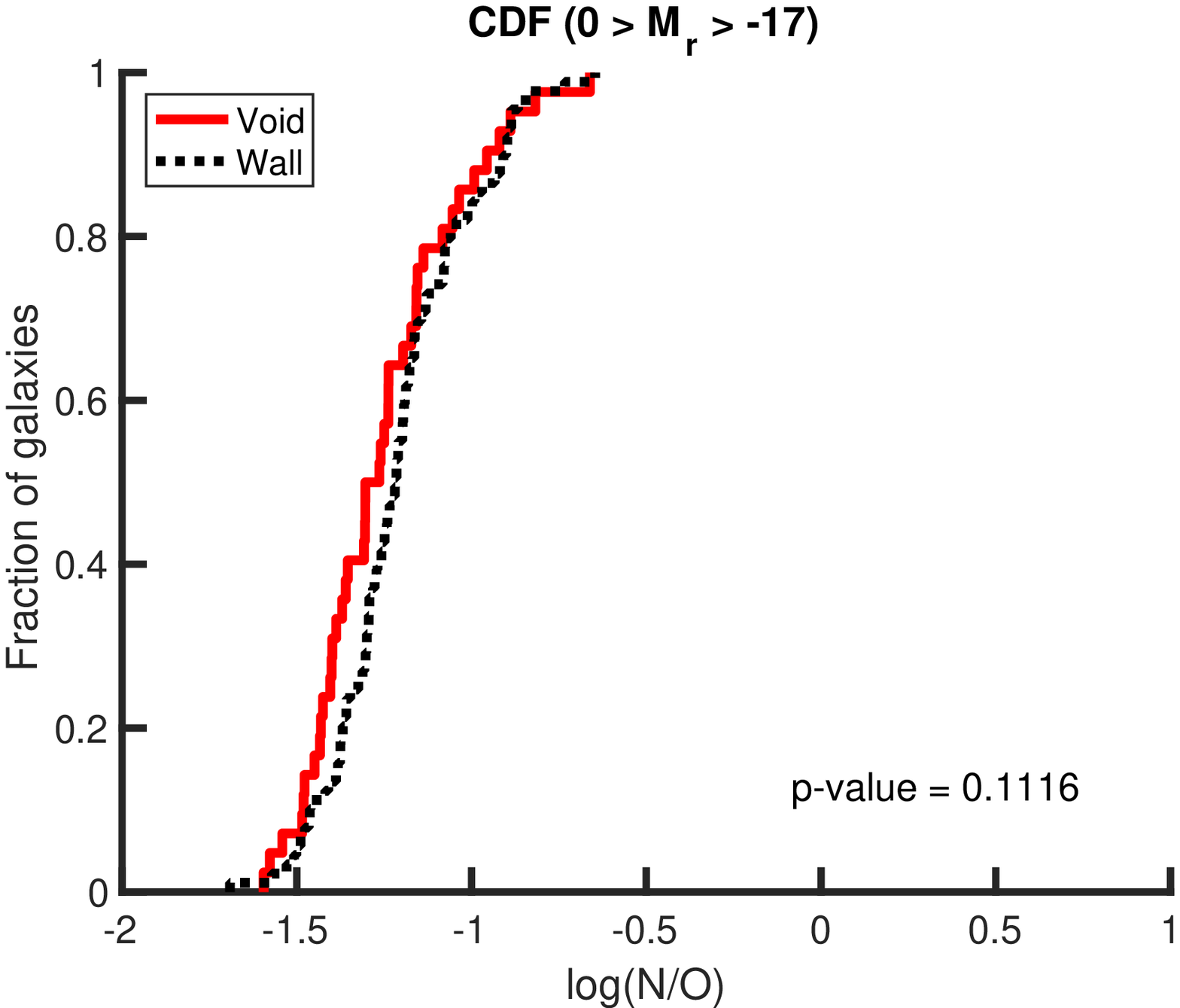}
    \caption{The ratio of nitrogen to oxygen of void dwarf (red solid line) and 
    wall dwarf (black dashed line) galaxies.  A two-sample KS test of the two 
    data sets results in an asymptotic $p$-value of 0.11, indicating an 11\% 
    probability that a test statistic greater than the observed value of 0.22 
    will be seen if the void sample was drawn from the wall sample.  This is 
    reflected visually, as the void galaxies appear to have a lower value of N/O 
    than the wall galaxies.  This is suggestive of a large-scale environmental 
    influence on the relative chemical abundances in dwarf galaxies.}
    \label{fig:NOratio}
\end{figure*}

In addition to studying the oxygen and nitrogen abundances relative to hydrogen, 
we also look at the ratio of nitrogen to oxygen.  The N/O abundance ratio 
suggests a slightly stronger environmental influence on the chemical evolution 
of dwarf galaxies than the oxygen and nitrogen abundances individually.  As can 
be seen in Fig. \ref{fig:NOratio}, there is a shift in the N/O ratio to lower 
values in the void dwarf galaxies than in the wall dwarf galaxies.  This 
difference is quantified in the KS test --- the test returned a probability of 
11.1\% that a test statistic greater than or equal to 0.22 will be measured if 
the void sample was drawn from the wall sample.  The void dwarf galaxies have 
lower N/O ratios by an average of $0.05\pm 0.074$ than the wall dwarf galaxies; 
the difference in the median values of the N/O ratio in the void and wall dwarf 
galaxy samples is 0.07.  However, like the shifts seen in the oxygen and 
nitrogen abundances, the shift in the N/O ratio for dwarf galaxies is not 
statistically significant.

\begin{figure}
    \centering
    \includegraphics[width=0.5\textwidth]{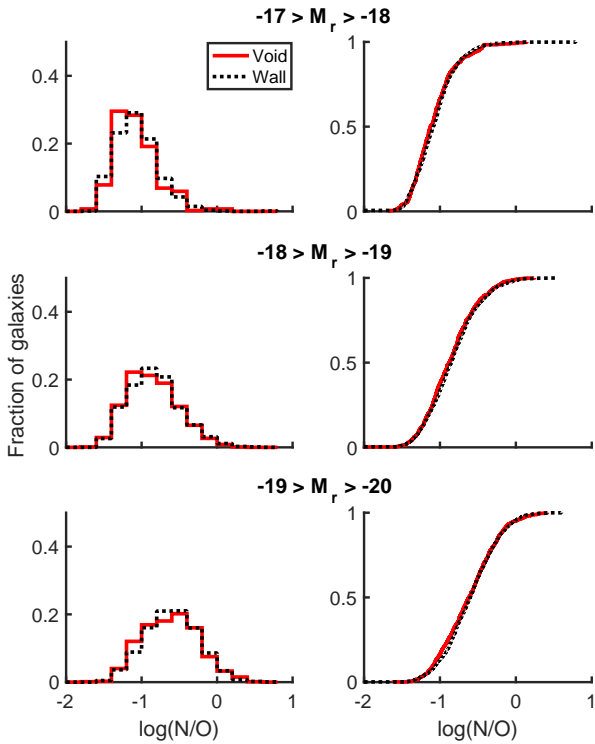}
    \caption{The ratio of nitrogen to oxygen of void (red solid line) and wall 
    (black dashed line) star-forming galaxies with $-17 > M_r > -18$ (top), 
    $-18 > M_r > -19$ (middle), and $-19 > M_r > -20$ (bottom).  The results of 
    a two-sample KS test of the two data sets in each absolute magnitude range 
    can be found in Table \ref{tab:stats}, in addition to other statistics of 
    the samples.  Both the histograms, CDFs, and statistics suggest a very 
    slight difference between the two populations, indicating a mild large-scale 
    environmental influence on the relative chemical abundances of star-forming 
    galaxies with absolute magnitudes $-17 > M_r > -20$.}
    \label{fig:NO_bright}
\end{figure}

We perform the same analysis with the N/O ratio on somewhat brighter galaxies, 
up through $M_r > -20$; the results of this analysis can be seen in Fig. 
\ref{fig:NO_bright} and in Table \ref{tab:stats}.  The shift towards lower N/O 
ratios for the void galaxies is small for all magnitude bins.  The direction of 
the shift between environments for the N/O ratio is consistent for all absolute 
magnitude bins: void galaxies have slightly lower N/O ratios than wall galaxies.  
This is only very weak evidence of a large-scale environmental influence on the 
relative abundances of elements in galaxies, but it is worth testing for in 
larger samples.

Fig. \ref{fig:NO_bright} indicates a shift towards higher values in the peak of 
the N/O distribution as the absolute magnitude of the galaxies increases.  There 
is a known positive correlation between the N/O ratio and the stellar mass of a 
galaxy, as discussed in Sec. \ref{sec:Mass_NO} below.  To test if this relation 
is causing the shift seen in Figs. \ref{fig:NOratio} and \ref{fig:NO_bright}, we 
downsampled the wall galaxies in each magnitude bin to match the void sample.  
The original shifts in the N/O ratio seen were still present after the 
downsampling; the observed shift in the N/O ratio is not due to any variations 
in the distribution of the stellar masses between the two environments.  In 
addition, if we are overestimating the temperatures in these galaxies due to an 
incorrect measurement of [\ion{O}{3}] $\lambda 4363$ (discussed above in Section 
\ref{sec:OH_NH}), that effect should cancel when we look at the ratio of 
nitrogen to oxygen.  This shift towards higher N/O values as a function of 
absolute magnitude indicates that brighter galaxies produce more nitrogen than 
fainter galaxies (relative to their oxygen abundance).  This result is consistent 
with the theory that nitrogen behaves as a secondary element in galaxies with 
high enough metallicity, if we assume a positive correlation between absolute 
magnitude and metallicity.

\floattable
\begin{deluxetable}{ccccccccc}
    \tablewidth{0pt}
    \tablecolumns{9}
    \tablecaption{Abundance statistics\label{tab:stats}}
    \tablehead{\colhead{Abs. mag. range} & \colhead{Environment} & \colhead{\# of galaxies} & \colhead{Average} & \colhead{Median} & \colhead{Average shift\tablenotemark{a}} & \colhead{Median shift\tablenotemark{a}} & \colhead{KS test statistic} & \colhead{$p$-value}}
    \startdata
        \cutinhead{$12 + \log (\text{O}/\text{H})$}\\
        \multirow{2}{*}{Dwarf galaxies} & Void & 42 & $7.99\pm 0.049$ & 8.04 & \multirow{2}{*}{$-0.07\pm 0.060$} & \multirow{2}{*}{-0.03} & \multirow{2}{*}{0.1322} & \multirow{2}{*}{0.6701}\\
         & Wall & 89 & $7.93\pm 0.036$ & 8.01 & & & & \\
        \multirow{2}{*}{$-17>M_r>-18$} & Void & 423 & $7.87\pm 0.016$ & 7.96 & \multirow{2}{*}{$0.01\pm 0.020$} & \multirow{2}{*}{-0.02} & \multirow{2}{*}{0.0455} & \multirow{2}{*}{0.5799}\\
         & Wall & 895 & $7.88\pm 0.012$ & 7.94 & & & & \\
        \multirow{2}{*}{$-18>M_r>-19$} & Void & 829 & $7.74\pm 0.013$ & 7.76 & \multirow{2}{*}{$-0.01\pm 0.016$} & \multirow{2}{*}{-0.04} & \multirow{2}{*}{0.0385} & \multirow{2}{*}{0.4000}\\
         & Wall & 1498 & $7.73\pm 0.009$ & 7.73 & & & & \\
        \multirow{2}{*}{$-19>M_r>-20$} & Void & 798 & $7.59\pm 0.013$ & 7.55 & \multirow{2}{*}{$-0.04\pm 0.017$} & \multirow{2}{*}{-0.05} & \multirow{2}{*}{0.0741} & \multirow{2}{*}{0.0062}\\
         & Wall & 1490 & $7.55\pm 0.010$ & 7.50 & & & & \\
         \cutinhead{$12 + \log (\text{N}/\text{H})$}\\
         \multirow{2}{*}{Dwarf galaxies} & Void & 42 & $6.74\pm 0.035$ & 6.77 & \multirow{2}{*}{$-0.02\pm 0.043$} & \multirow{2}{*}{-0.01} & \multirow{2}{*}{0.1129} & \multirow{2}{*}{0.8383}\\
         & Wall & 89 & $6.72\pm 0.025$ & 6.75 & & & & \\
        \multirow{2}{*}{$-17>M_r>-18$} & Void & 423 & $6.80\pm 0.011$ & 6.82 & \multirow{2}{*}{$0.00\pm 0.014$} & \multirow{2}{*}{-0.01} & \multirow{2}{*}{0.0401} & \multirow{2}{*}{0.7364}\\
         & Wall & 895 & $6.80\pm 0.008$ & 6.81 & & & & \\
        \multirow{2}{*}{$-18>M_r>-19$} & Void & 829 & $6.87\pm 0.009$ & 6.85 & \multirow{2}{*}{$0.01\pm 0.011$} & \multirow{2}{*}{0.02} & \multirow{2}{*}{0.0538} & \multirow{2}{*}{0.0873}\\
         & Wall & 1498 & $6.89\pm 0.007$ & 6.87 & & & & \\
        \multirow{2}{*}{$-19>M_r>-20$} & Void & 798 & $6.97\pm 0.009$ & 6.94 & \multirow{2}{*}{$-0.02\pm 0.011$} & \multirow{2}{*}{-0.02} & \multirow{2}{*}{0.0455} & \multirow{2}{*}{0.2272}\\
         & Wall & 1490 & $6.95\pm 0.007$ & 6.93 & & & & \\
         \cutinhead{$\log (\text{N}/\text{O})$}\\
          \multirow{2}{*}{Dwarf galaxies} & Void & 42 & $-1.25\pm 0.060$ & -1.28 & \multirow{2}{*}{$0.05\pm 0.074$} & \multirow{2}{*}{0.07} & \multirow{2}{*}{0.2191} & \multirow{2}{*}{0.1116}\\
         & Wall & 89 & $-1.21\pm 0.044$ & -1.22 & & & & \\
        \multirow{2}{*}{$-17>M_r>-18$} & Void & 423 & $-1.08\pm 0.020$ & -1.13 & \multirow{2}{*}{$-0.00\pm 0.024$} & \multirow{2}{*}{0.04} & \multirow{2}{*}{0.0588} & \multirow{2}{*}{0.2657}\\
         & Wall & 895 & $-1.08\pm 0.014$ & -1.09 & & & & \\
        \multirow{2}{*}{$-18>M_r>-19$} & Void & 829 & $-0.86\pm 0.015$ & -0.88 & \multirow{2}{*}{$0.02\pm 0.019$} & \multirow{2}{*}{0.03} & \multirow{2}{*}{0.0550} & \multirow{2}{*}{0.0763}\\
         & Wall & 1498 & $-0.84\pm 0.012$ & -0.85 & & & & \\
        \multirow{2}{*}{$-19>M_r>-20$} & Void & 798 & $-0.62\pm 0.016$ & -0.62 & \multirow{2}{*}{$0.02\pm 0.020$} & \multirow{2}{*}{0.02} & \multirow{2}{*}{0.0564} & \multirow{2}{*}{0.0715}\\
         & Wall & 1490 & $-0.60\pm 0.012$ & -0.60 & & & & \\
    \enddata
    \tablecomments{Statistics of the gas-phase oxygen, nitrogen, and nitrogen 
    relative to oxygen abundances in void and wall galaxies in each of the 
    absolute magnitude ranges listed.  Most of these results are not 
    statistically significant, as shown in Figs. 
    \ref{fig:met1sig}--\ref{fig:NO_bright}.  However, the shifts in chemical 
    abundances between the two environments are predominately in the same 
    direction for each of the magnitude bins, suggesting that there is some 
    influence on the chemical evolution of galaxies by the large-scale 
    environment.   Void galaxies have slightly higher oxygen and nitrogen 
    abundances than wall galaxies, but void galaxies have slightly lower N/O 
    ratios than wall galaxies.}
    \tablenotetext{a}{Wall -- Void (Positive shifts indicate that the wall 
    values are greater than the void values; negative shifts indicate that the 
    void values are greater than the wall values.)}
\end{deluxetable}

\subsection{N/O v. O/H}

Comparing the N/O ratio with the gas-phase oxygen abundance in a galaxy can help 
us understand the nucleosynthesis of nitrogen in galaxies.  When the metallicity 
of a galaxy is low, stars created from this gas do not have enough carbon to 
efficiently produce helium via the CNO cycle.  As a result, any nitrogen 
produced in these stars will behave as a primary element --- it will be produced 
in the same relative quantity as oxygen.  However, when the metallicity of a 
galaxy is high enough, stars are created with enough seed carbon to initiate the 
CNO cycle at an earlier stage in the star's life.  As a result, nitrogen will 
behave as a secondary element and will be produced in a larger quantity relative 
to the primary elements (like oxygen and carbon).  By studying the relation 
between N/O and the metallicity of a galaxy, we should be able to discern the 
critical metallicity at which nitrogen switches from a primary to a secondary 
element.

\begin{figure}
    \includegraphics[width=0.5\textwidth]{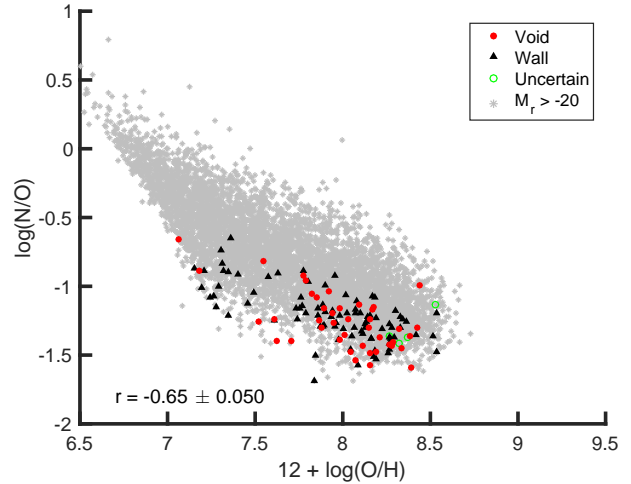}
    \caption{N/O as a function of O/H for star-forming void (red circles) and 
    wall (black triangles) dwarf galaxies.  Error bars have been omitted for 
    clarity.  N/O is expected to be constant for 
    $12 + \log (\text{O}/\text{H}) \lesssim 8.5$, as the metallicity of a 
    galaxy's ISM is too low for stars to be created with enough of the heavy 
    elements to undergo the CNO cycle early enough in their lifetimes.  As a 
    result, nitrogen behaves as a primary element at galactic metallicities less 
    than approximately 8.5.  To place the dwarf galaxies in the context of the 
    general galaxy population, we also plot (gray stars) all star-forming 
    galaxies fainter than $M_r > -20$.}
    \label{fig:OvN}
\end{figure}

Our results for N/O v. metallicity can be seen in Fig. \ref{fig:OvN}.  Unlike 
many previous comparisons of N/O and metallicity 
\citep[for example,][]{VilaCostas93, Thuan95, Henry00, Pilyugin02, Lee04, 
Pilyugin04, Nava06, vanZee06a, PerezMontero09, Amorin10, Berg12}, we do not find 
a constant value for N/O as a function of O/H for dwarf galaxies (nor for any of 
the somewhat brighter galaxies).  \cite{Shields91, Contini02, Nicholls14b} also 
find little or no evidence of a plateau in their study.  Instead of a constant 
value for N/O as a function of O/H at low metallicities, we find a slight 
decrease in the N/O ratio as the metallicity increases; a linear fit to the 
dwarf galaxies reveals a slope of $-0.38\pm 0.078$.  This is close to the 
footnoted results of \cite{Andrews13}, who find a slope of $-0.21$ for their 
stellar mass-binned galaxies with metallicities 
$12 + \log (\text{O}/\text{H}) < 8.5$.  The average value of 
$\log(\text{N}/\text{O})$ for the void dwarf galaxies is $-1.25\pm 0.060$ with a 
median value of $-1.28$, while the average value for the wall dwarf galaxies is 
$-1.21\pm 0.044$ with a median of $-1.22$.  As shown in Fig. \ref{fig:NOratio}, 
the void dwarf galaxies have slightly less nitrogen relative to oxygen than do 
dwarf galaxies in more dense regions.  Both these median values are higher than 
that of \cite{Andrews13}, and these average values are higher than that of 
\cite{Izotov99} and \cite{Nava06}.  

\begin{figure}
    \centering
    \includegraphics[width=0.5\textwidth]{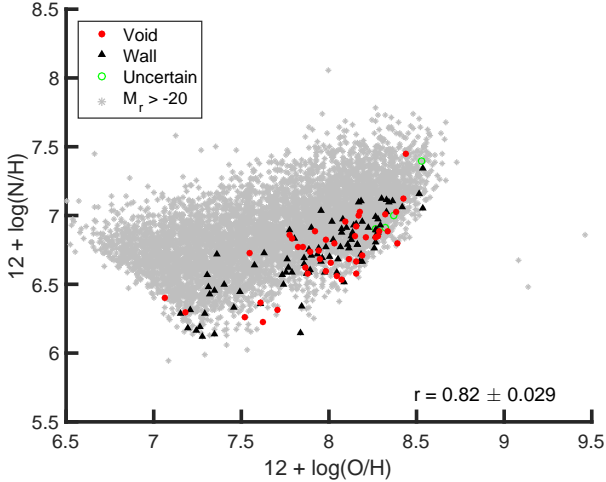}
    \caption{N/H as a function of O/H for star-forming void (red circles) and 
    wall (black triangles) dwarf galaxies.  Error bars have been omitted for 
    clarity.  There is a positive correlation between the two abundances.  With 
    a best-fit slope less than 1, we see that the synthesis of nitrogen in these 
    galaxies is primary.}
    \label{fig:OHvNH}
\end{figure}

If a plateau in the O/H--N/O relation exists, then we should see a slope of 1 in 
the O/H--N/H relation.  When looking at N/H as a function of O/H in Fig. 
\ref{fig:OHvNH}, we see that there is a correlation between the nitrogen and 
oxygen abundances.  However, a best fit to the dwarf galaxies reveals a slope of 
only $0.62\pm 0.078$ --- the nitrogen abundance increases at a slower rate than 
the oxygen abundance.  This result matches the negative relationship between the 
metallicity and the N/O ratio seen in Fig. \ref{fig:OvN}.  If we examine only 
the low metallicity ($12 + \log(\text{O}/\text{H}) < 7.6$) star-forming galaxies 
with $M_r > -20$, a linear fit produces a slope of $0.05\pm 0.019$ in Fig. 
\ref{fig:OHvNH} and a slope of $-0.94\pm 0.019$ in Fig. \ref{fig:OvN}.  This is 
in sharp contrast to the star-forming galaxies with $M_r > -20$ that have 
metallicities $12 + \log(\text{O}/\text{H}) > 7.6$, where their slope in Fig. 
\ref{fig:OHvNH} is $0.60\pm 0.022$ and $-0.39\pm 0.023$ in Fig. \ref{fig:OvN}.  
It appears that the nitrogen production is independent of the amount of oxygen 
produced in low metallicity systems.  At normal metallicities 
($7.6 < 12 + \log(\text{O}/\text{H}) < 8.5$), there exists a positive 
relationship between the production of nitrogen and oxygen, although the ratio 
of N/O produced depends on the galaxy's metallicity.

There is no difference between void and wall galaxies in the relationship of 
oxygen and nitrogen production in the low metallicity sample.  There is a slight 
difference in slopes between the void and wall galaxies with normal 
metallicities, where the void galaxies have a larger slope in the relationship 
between O/H and N/H and a smaller slope in the relationship between O/H and N/O.  
While statistically significant, the difference in the slopes between the two 
environments is not large enough to be physically relevant.  The significant 
scatter in both Figs. \ref{fig:OHvNH} and \ref{fig:OvN} indicates that the 
described relationships between the production of nitrogen and oxygen are only 
global trends in the nucleosynthesis of the galaxies.

\subsection{Mass--N/O relation}\label{sec:Mass_NO}

\begin{figure}
    \includegraphics[width=0.5\textwidth]{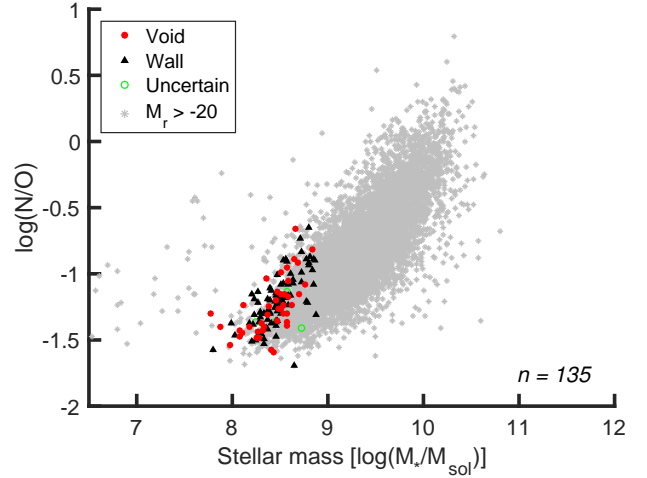}
    \caption{Mass--N/O relation for star-forming void (red circles) and wall 
    (black triangles) dwarf galaxies.  Error bars have been omitted for clarity.  
    N/O is expected to remain constant for low stellar masses and increase 
    steadily for larger masses, due to the mass-metallicity relation and the 
    primary v. secondary synthesis of nitrogen.  To place the dwarf galaxies in 
    the context of the general galaxy population, we also plot (gray stars) the 
    star-forming galaxies with $M_r > -20$.}
    \label{fig:MNO}
\end{figure}

Just as there is a well-known mass--metallicity relation for galaxies 
\citep[where the metallicity increases with stellar mass; see][for example]{Tremonti04}, 
there is also a mass--N/O relation.  We expect to see a primary N/O plateau in 
the mass--N/O relation, since galaxies with lower stellar masses have not yet 
produced enough heavy elements to synthesize more nitrogen than oxygen.  Beyond 
the low-mass limit, there should be a steady increase in the N/O ratio as a 
function of stellar mass, due to secondary nitrogen enrichment.  Our dwarf 
galaxies in Fig. \ref{fig:MNO} show a steady increase in N/O as a function of 
stellar mass; there is a hint of the beginnings of a plateau for 
$\log(M_*/M_{\odot}) \lesssim 8$.  The lack of a plateau here could be a result 
of our limited stellar mass range for the dwarf galaxies.  A linear fit to our 
dwarf galaxies reveals a slope of $0.6\pm 0.12$, which is much stronger than the 
slope of $0.30$ found by \cite{Andrews13}.

From Fig. \ref{fig:MNO}, we conclude that the N/O plateau, if one exists, starts 
around $\log(M_*/M_{\odot}) \approx 8$.  This is at a much lower mass than that 
found by \cite{Andrews13} --- they claim the N/O plateau exists for galaxies 
with $\log(M_*/M_{\odot}) < 8.9$.  However, our relationship between stellar 
mass and N/O matches Fig. 3 of \cite{Amorin10} as well as the results of 
\cite{PerezMontero09} and \cite{PerezMontero13}.  None of our data samples 
display an obvious N/O plateau above a stellar mass 
$\log(M_*/M_{\odot}) > 8$, indicative of primary nitrogen versus secondary 
nitrogen production in galaxies.  This could indicate that the switch from 
primary to secondary nitrogen production occurs at a much lower stellar mass 
than found by \cite{Andrews13}.  More low-mass galaxies are needed to extend 
this relation below $\log(M_*/M_{\odot}) < 8$.

\subsection{Color--N/O relation}

\begin{figure*}
    \includegraphics[width=0.49\textwidth]{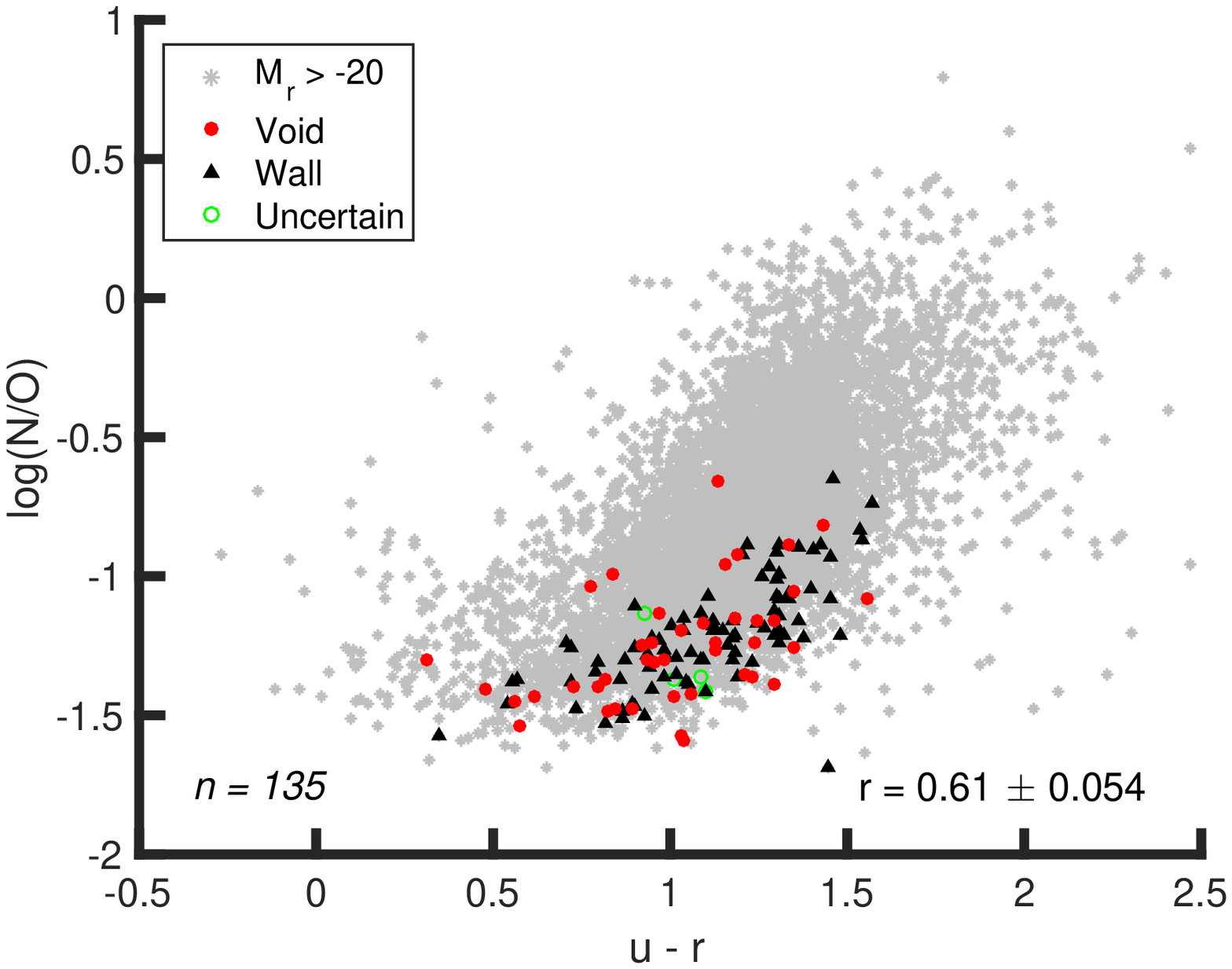}
    \includegraphics[width=0.49\textwidth]{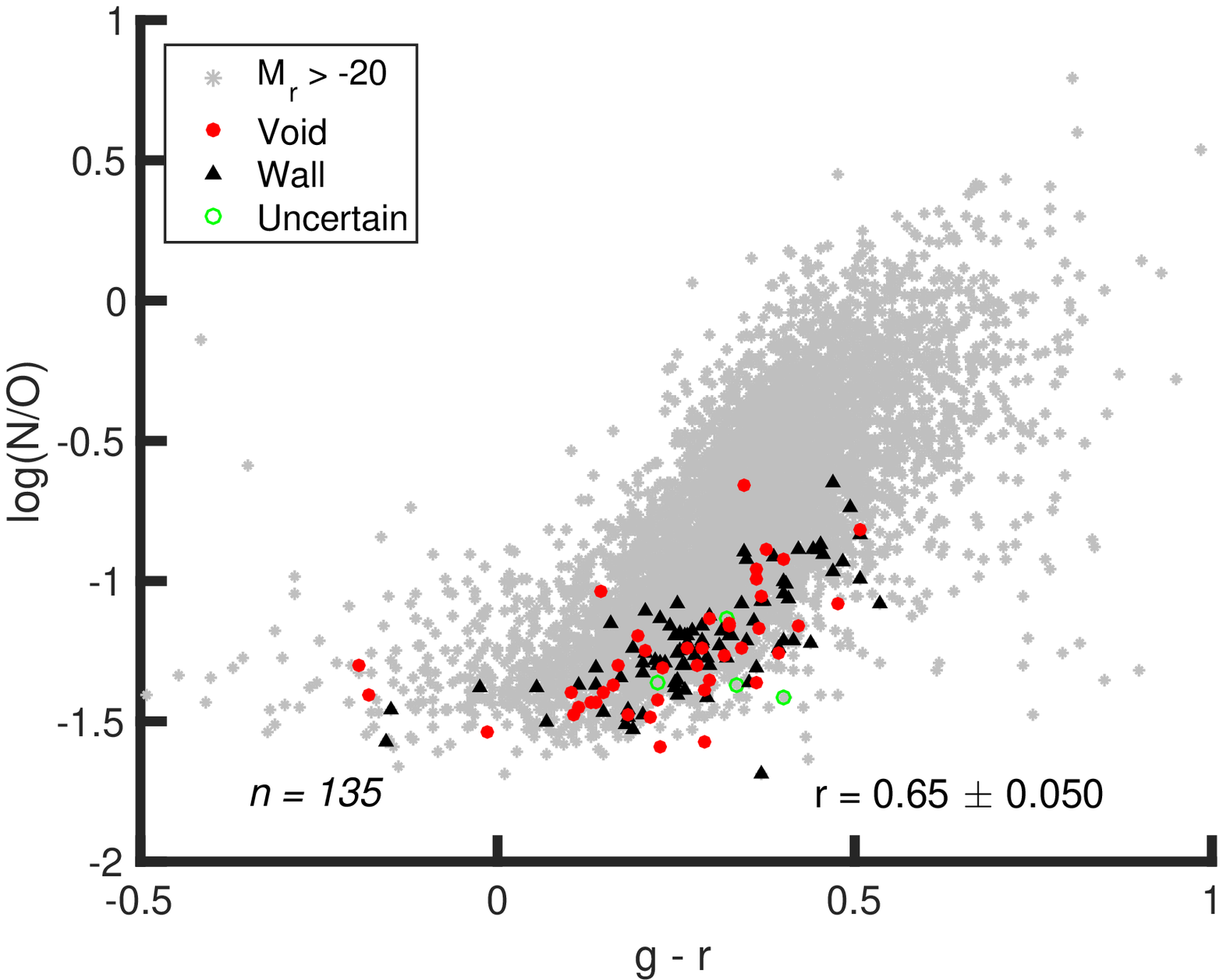}
    \caption{Color ($u-r$ and $g-r$) vs. the N/O ratio for star-forming void 
    (red circles) and wall (black triangles) dwarf galaxies.  Error bars have 
    been omitted for clarity.  N/O is expected to increase as galaxies become 
    redder if there is a time delay between the release of oxygen and nitrogen.  
    To place the dwarf galaxies in context, we also plot the star-forming 
    galaxies with $M_r > -20$ (gray stars).}
    \label{fig:color_NO}
\end{figure*}

As \cite{vanZee06a,Berg12} discuss, a time delay between the release of nitrogen 
and oxygen will result in a positive relationship between the N/O ratio and the 
color of a galaxy.  If oxygen is primarily produced in higher mass stars (and 
since these stars die earlier than the intermediate-mass stars responsible for 
the synthesis of nitrogen), then, for a given star formation episode, the oxygen 
produced will be released on a shorter time scale than the nitrogen.  As a 
result, the amount of nitrogen relative to oxygen should increase as the hotter, 
more massive stars die and the galaxy becomes redder.  This is exactly what we 
see in Fig. \ref{fig:color_NO}.  We use rest-frame colors K-corrected to a 
redshift of 0.1; they are corrected for galactic extinction and calculated with 
model magnitudes \citep{Choi10}.  \cite{vanZee06a, Berg12} also found an 
increase in the N/O ratio as a function of color.

\subsection{(s)SFR--N/O relation}

\begin{figure*}
    \includegraphics[width=0.49\textwidth]{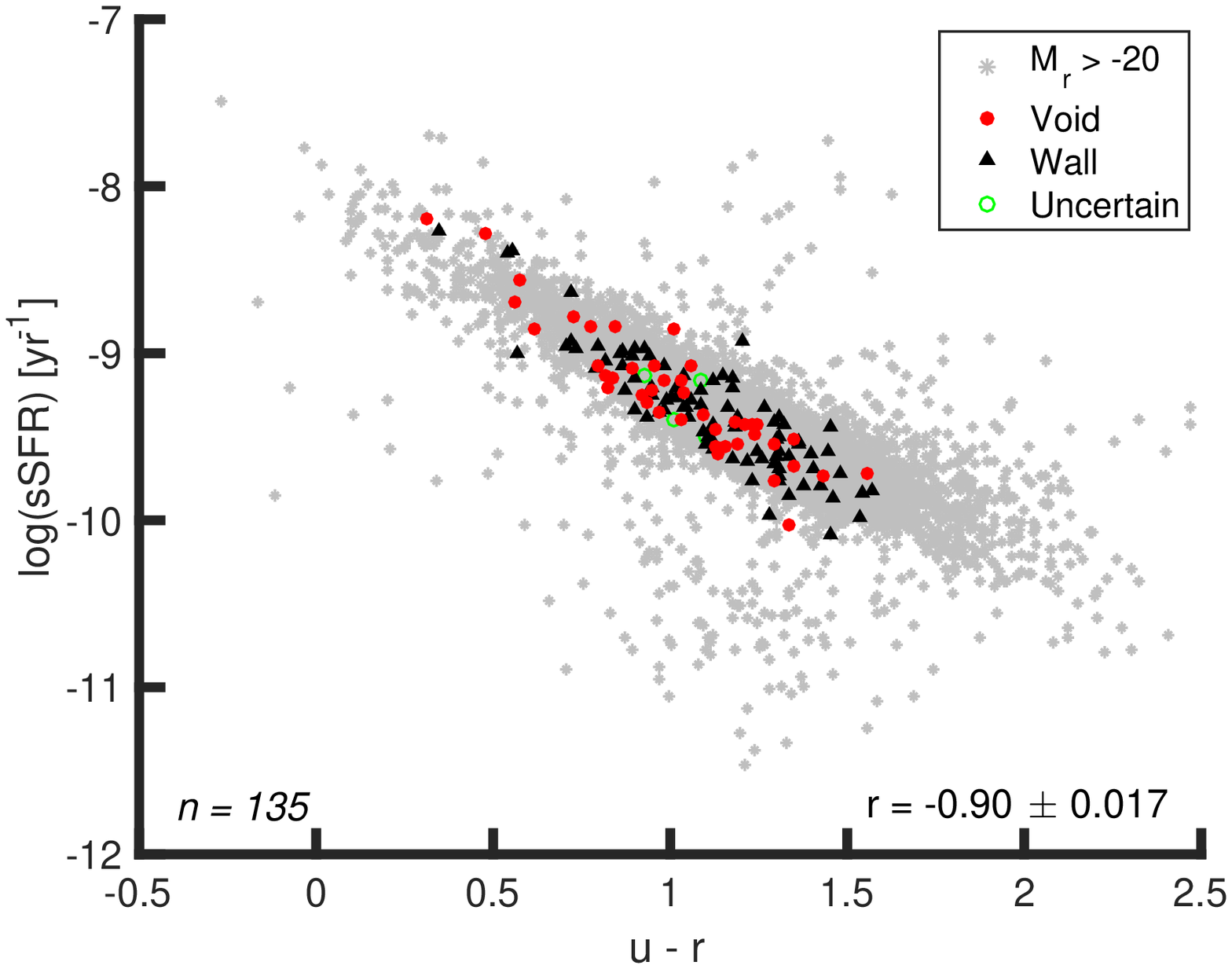}
    \includegraphics[width=0.49\textwidth]{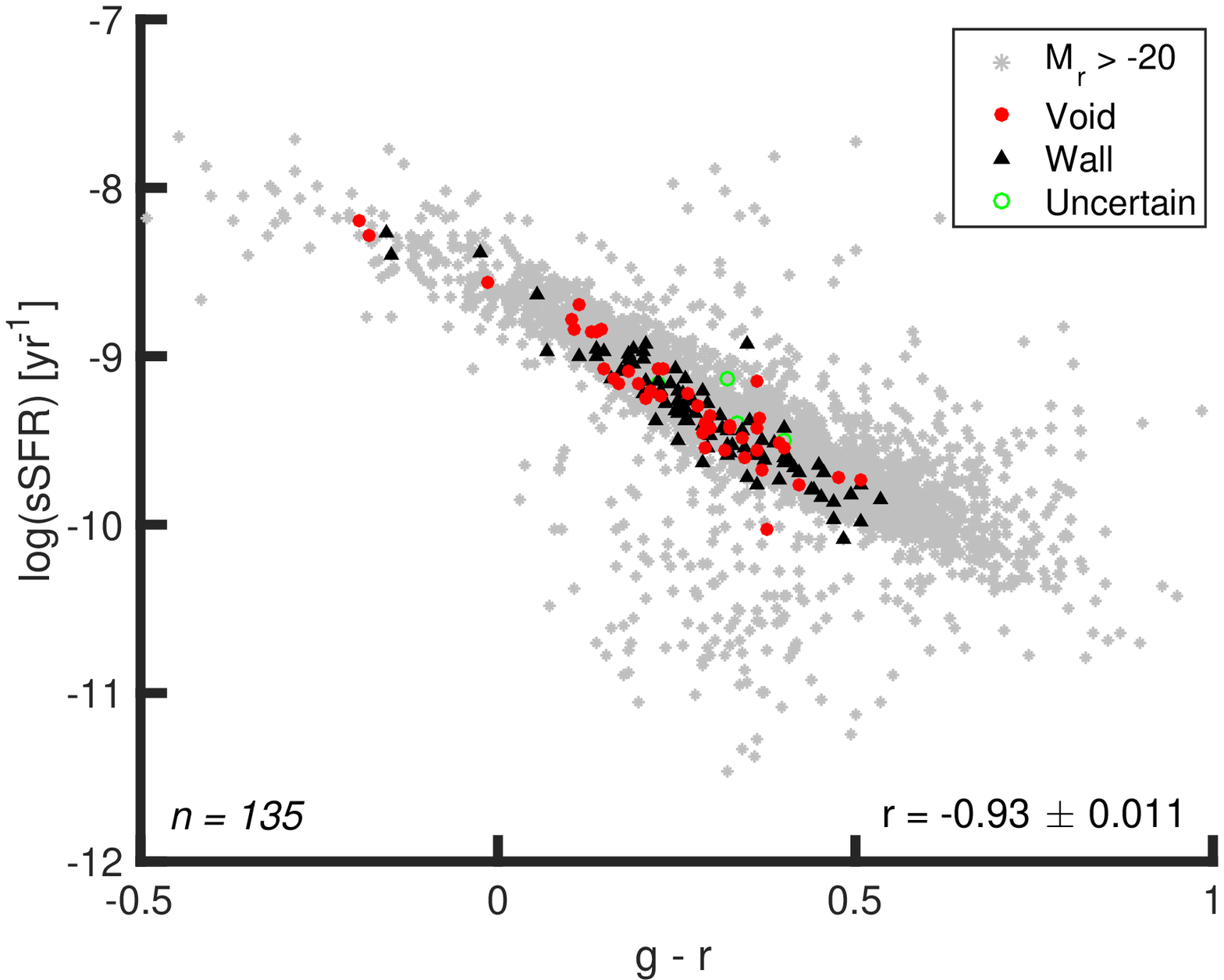}
    \caption{Color ($u-r$ and $g-r$) vs. sSFR for star-forming void (red 
    circles) and wall (black triangles) dwarf galaxies.  It is obvious that 
    there is a negative correlation between the $\log (sSFR)$ and the color of a 
    galaxy.  To place the dwarf galaxies in context, we also show the 
    star-forming galaxies with $M_r > -20$ (gray stars).}
    \label{fig:color_sSFR}
\end{figure*}

\begin{figure*}
    \includegraphics[width=0.49\textwidth]{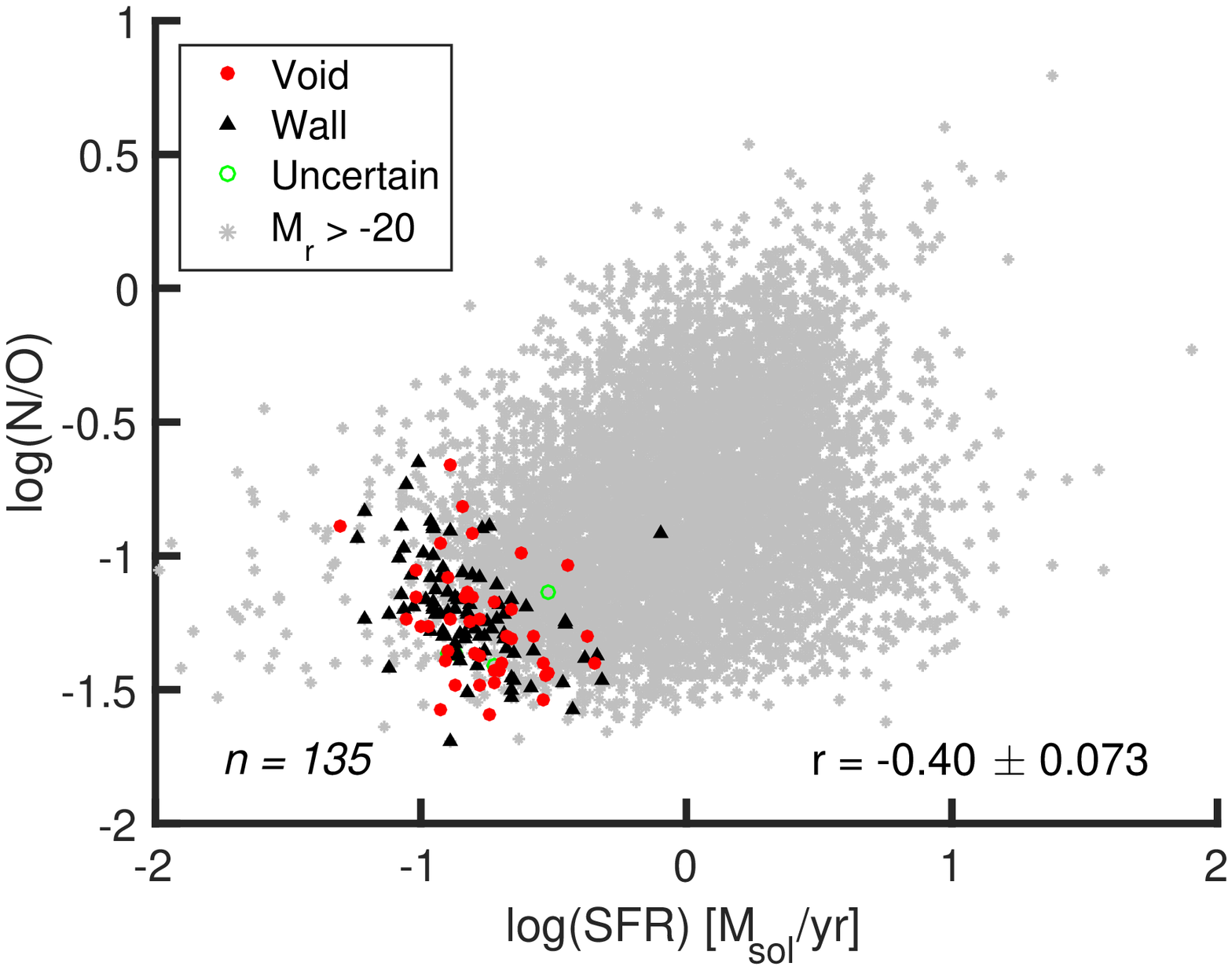}
    \includegraphics[width=0.49\textwidth]{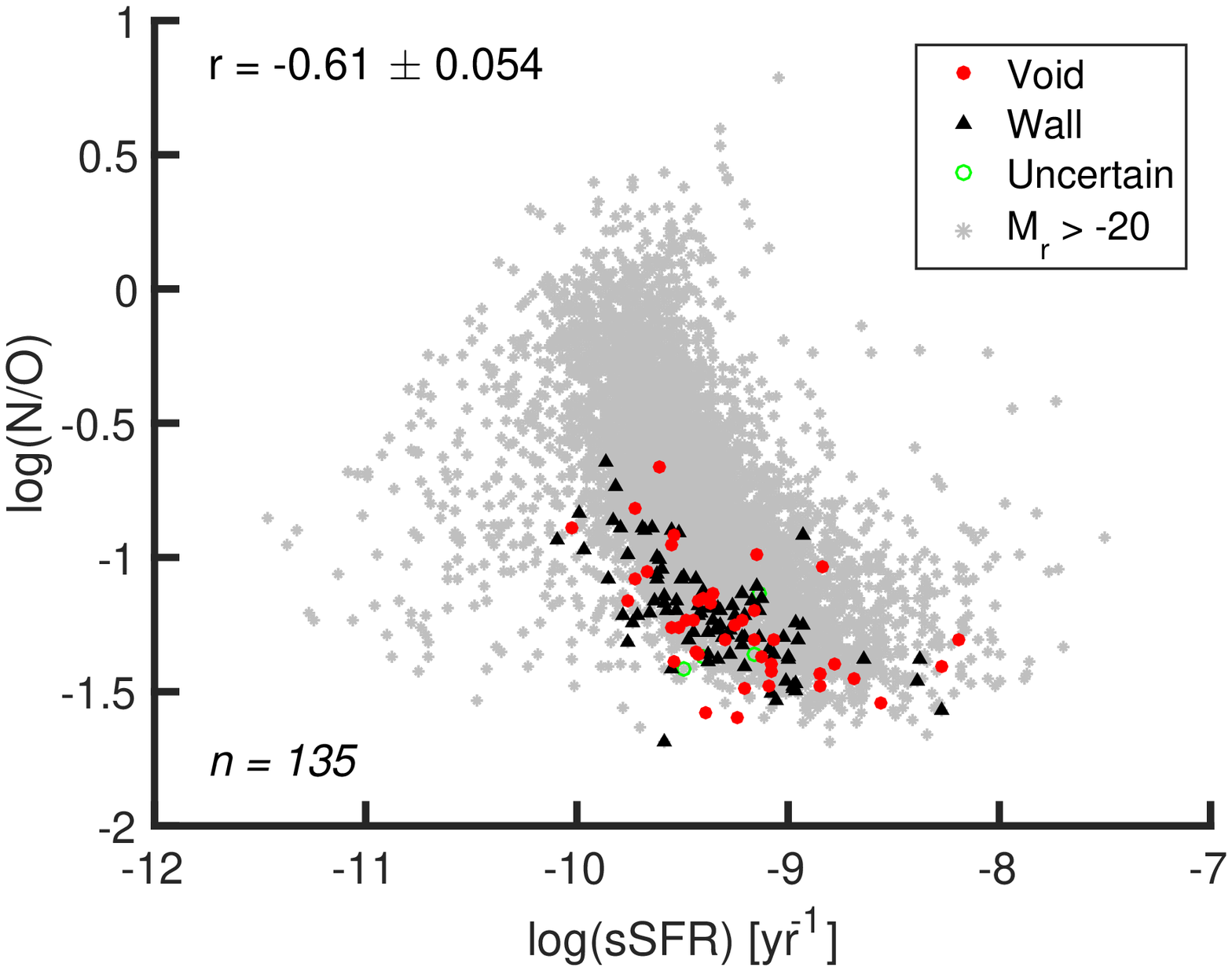}
    \caption{SFR--N/O and sSFR--N/O relations for star-forming void 
    (red circles) and wall (black triangles) dwarf galaxies.  Error bars have 
    been omitted for clarity.  Based on the correlations between SFR and stellar 
    mass and sSFR and color, we expect the N/O ratio to increase with increasing 
    SFR and decrease with sSFR.  To place the dwarf galaxies in context, we also 
    show the star-forming galaxies with $M_r > -20$ (gray stars).}
    \label{fig:SFR_NO}
\end{figure*}

We expect there to be a correlation between star formation rate (SFR) or 
specific (per unit stellar mass) star formation rate (sSFR) and the N/O ratio in 
galaxies as a result of the positive correlations between the SFR and stellar 
mass of a galaxy \citep{Brinchmann04} and the sSFR and the color of a galaxy 
(Fig. \ref{fig:color_sSFR}).  As shown in Fig. \ref{fig:MNO}, the N/O ratio 
increases with increasing stellar mass.  As a result, we expect there to be a 
positive correlation between the SFR and N/O ratio in our galaxies.  This can be 
seen in the sample of star-forming galaxies with $M_r > -20$ in the left panel 
of Fig. \ref{fig:SFR_NO}.  Due to the large scatter in this relation, the blue 
star-forming dwarf galaxies exhibit a negative correlation between their N/O 
ratios and their SFRs.  These SFR are aperture-corrected to estimate the total 
SFR in the galaxy (not just within the SDSS fiber).

The N/O ratio is expected to decrease as sSFR increases in galaxies, as is shown 
in the right panel of Fig. \ref{fig:SFR_NO}.  Bluer galaxies have higher sSFR.  
Fig. \ref{fig:color_NO} shows that there is a positive correlation between the 
color and the N/O ratio, such that bluer galaxies have lower N/O ratios.  As a 
result, we are not surprised to see that the N/O ratio decreases as the sSFR 
increases.

%
%
\section{Large-scale environmental influence}\label{sec:environment}

The majority of the shifts in the gas-phase abundances of oxygen, nitrogen, and 
the N/O ratio seen in galaxies fainter than $L_*$ are small and statistically 
insignificant.  However, they occur in almost the same direction across all 
magnitude bins.  This trend suggests that the gas-phase abundances may be 
influenced by the galaxies' large-scale environments.  \cite{Shields91} find no 
offset in the N/O ratio between cluster and field galaxies, despite the 
difference in O/H they observe.  Similar to us, \cite{Contini02, Pilyugin02} 
also find a statistically insignificant shift in the N/O ratio of cluster 
galaxies, although they find that these galaxies have lower N/O ratios than 
field spiral galaxies.  Based on Fig. \ref{fig:NOratio} and the statistics in 
Table \ref{tab:stats}, we find weak evidence that void dwarf galaxies have a 
smaller N/O ratio than dwarf galaxies in more dense regions.  This means that 
void dwarf galaxies may have more oxygen than wall dwarf galaxies, and/or void 
dwarf galaxies could have less nitrogen than wall dwarf galaxies.  Here we 
discuss these possibilities and explore their implications for the large-scale 
environmental impact on the formation and evolution of galaxies.

Table \ref{tab:stats} suggests a slight large-scale environmental dependence of 
the oxygen and nitrogen abundances (relative to hydrogen), where void galaxies 
have slightly more O/H and N/H than wall galaxies.  This small difference is 
not apparent when looking at Figs. \ref{fig:met1sig} and \ref{fig:N_1sig}.  
However, the N/O ratio amplifies this large-scale environmental effect so that a 
shift in the mean (or median) of the two populations can be seen in Fig. 
\ref{fig:NOratio}.  We hesitate to combine the results across all magnitude bins 
in an effort to improve their significance.  Instead, we look towards the future 
to analyze a larger sample of galaxies to increase the statistical significance 
of these results.

\subsection{Higher metallicities in void dwarf galaxies}

A slightly higher metallicity in void dwarf galaxies than wall dwarf galaxies 
may be evidence of a difference in the dark matter halo mass to stellar mass 
ratio between the two environments.  Simulations by \cite{Jung14} and 
\cite{Tonnesen15} have shown that the dark matter halo masses of void central 
galaxies are larger than that of wall central galaxies for a given stellar mass.  
Due to their environment, void dwarf galaxies are more likely to be in the 
center of their own dark matter halo.  Wall dwarf galaxies, on the other hand, 
are much more likely to be a satellite galaxy within a much larger dark matter 
halo; the simulation results mentioned above would not apply to these wall dwarf 
galaxies.  However, because the wall dwarf galaxies studied here have 
sufficiently high sSFRs, they most likely live in a small-scale environment very 
similar to that of the void dwarf galaxies, as discussed in Paper 1.  As a 
result, it is likely that (and should be tested to see if) the wall dwarf 
galaxies in this study are actually central galaxies.  

Applying the results of these simulations to our dwarf galaxy sample, if the 
ratio of dark matter halo mass to stellar mass is larger in void galaxies, the 
metals ejected from a void galaxy's ISM into its circumgalactic medium are more 
likely to fall back onto the ISM than in a wall galaxy with the same stellar 
mass, since the void galaxy's virial radius and potential well are larger.  As a 
result, two dwarf galaxies with the same stellar mass in these two different 
large-scale environments can have different metallicities --- void dwarf 
galaxies would have higher metallicities than wall dwarf galaxies, matching what 
we see in Table \ref{tab:stats}.

\subsection{Lower N/O ratios in void dwarf galaxies}

A difference in the N/O ratio between void dwarf galaxies and wall dwarf 
galaxies could be a result of the difference in the synthesis of nitrogen in 
galaxies within these two large-scale environments.  If void galaxies are 
retarded in their star formation (as simulations of the $\Lambda$CDM cosmology 
suggest), then cosmic downsizing would reduce the star formation rate at late 
times much more in wall galaxies than in void galaxies.  As a result, the 
minimum gas-phase metallicity required for the production of secondary nitrogen 
in walls would be achieved at an earlier time in the galaxy's lifetime than in a 
void galaxy.  This would cause the N/O ratio in wall galaxies to be larger than 
that in voids.  \cite{vanZee06a} suggest that a galaxy with a declining star 
formation rate (wall galaxies) will have a higher nitrogen-to-oxygen yield than 
a galaxy with a constant star formation rate (void galaxies).  This is due to 
more oxygen being released into the ISM as a result of the ongoing star 
formation in the void galaxies.  This explanation is supported by the color-N/O 
diagram in Fig. \ref{fig:color_NO}, which reveals that redder galaxies have 
higher N/O ratios.  The correlation between color and the N/O ratio found in 
\cite{vanZee06a, Berg12}, and Fig. \ref{fig:color_NO} is a result of declining 
star formation rates \citep{vanZee06a}.  Therefore, the shift in the N/O ratio 
we see between void and wall galaxies may be observational evidence of retarded 
star formation in void galaxies as a result of cosmic downsizing.

Another explanation that would lead to a shift in the N/O ratio between 
environments is a difference in the ratio of intermediate and high mass stars 
produced in void and wall dwarf galaxies.  For there to be more oxygen relative 
to nitrogen in void dwarf galaxies, the percent of higher mass stars produced in 
a star formation episode would be higher than that in wall dwarf galaxies.  This 
would indicate a varying initial mass function (IMF) as a function of 
large-scale environment.  Previous studies have been inconclusive when testing 
for a varying IMF \citep[see][for example]{Kroupa01, Kroupa02, Hoversten08, 
Meurer09}.  It is beyond the scope of this paper to elaborate on this 
explanation.

\subsection{N/O ratio for extreme low-metallicity galaxies}

While Paper 1 shows that there is not a special population of extremely 
low-metallicity dwarf galaxies residing in voids, we want to look in particular 
at the N/O ratio of extreme low-metallicity galaxies.  For the 21 dwarf galaxies 
with $12 + \log (\text{O}/\text{H}) < 7.6$ identified in Paper 1, we see from 
Fig. \ref{fig:OHvNH} that their N/H ratios are also some of the lowest in the 
dwarf galaxy sample.  However, as shown in Fig. \ref{fig:OvN}, their N/O ratios 
cover the range of N/O ratio values of all the dwarf galaxies studied.  This is 
consistent with the expectation that nitrogen behaves as a primary element for 
galaxies with low and moderate metallicities.

\floattable
\input{t2}

%
%
\section{Conclusions}

The nucleosynthesis of nitrogen is a vital component of the chemical evolution 
of galaxies in our Universe.  We estimate the nitrogen abundance and N/O ratio 
of dwarf galaxies using the direct $T_e$ method and spectroscopic line flux 
measurements from the SDSS DR7 sample as re-processed in the MPA-JHU catalog.  
The 135 galaxies analyzed suggest a slight large-scale environmental dependence 
of the N/O ratio, where void dwarf galaxies could have a lower N/O ratio than 
dwarf galaxies in more dense environments.  Thus, the large-scale 
($\sim 10\text{ Mpc}$) environment might influence the chemical evolution of 
dwarf galaxies.

We find small, statistically insignificant shifts in the mean (or median) N/O 
ratio for galaxies between the void and more dense regions across all blue, 
star-forming galaxies with $M_r > -20$.  These shifts are somewhat more 
significant, however, when we look at the entire sample of galaxies.  Each 
magnitude bin is shifted in the same direction and are potentially very 
interesting, as they might indicate delayed star formation histories, more 
constant star formation rates, and larger dark-matter-halo-mass to stellar mass 
ratios in void galaxies, as discussed in Section \ref{sec:environment}.  A 
larger sample would help test these results.  We look to increase the sample and 
probe a larger magnitude (and mass) range of dwarf galaxies in Douglass \& 
Vogeley (2017, in preparation).

In addition, we look at the relationship between the N/O ratio and other 
physical characteristics of our dwarf galaxies.  In the relation between N/O and 
O/H, our galaxies all reside on the so-called ``nitrogen plateau,'' where the 
N/O ratio is predicted to be constant for low- and intermediate-metallicities.  
However, instead of a constant value for these galaxies, we find a negative 
correlation between the N/O and O/H ratios.  Our dwarf galaxies show a positive 
correlation between stellar mass and the N/O ratio.  These dwarf galaxies have 
some of the lowest N/O ratios for both their color and (s)SFR.  Beyond the 
suggestive large-scale environmental dependence of the N/O ratio, there is no 
clear large-scale environmental dependence in any of these relationships.

The N/O ratios of the extremely metal-poor dwarf galaxies are no different than 
that of the remaining dwarf galaxy sample, though their N/H abundance is also 
extremely low.  A more detailed study of these 21 extremely metal-poor dwarf 
galaxies is recommended to confirm their abundance values and discover any 
characteristics shared by the population.

Although SDSS provides spectroscopic observations for over 800,000 galaxies, 
only 135 are dwarf galaxies with emission line fluxes necessary to estimate the 
gas-phase chemical abundances using the direct $T_e$ method.  The greatest 
limiting factor in this sample is the requirement of the [\ion{O}{2}] 
$\lambda 3727$ doublet in the abundance calculations.  We seek to develop a 
work-around for this emission line in Douglass \& Vogeley (2017, in preparation) 
to greatly increase our sample of dwarf galaxies with abundance estimates.  
These estimated ionic abundances can then be compared with environmental 
dependence of star formation and abundance predictions from high-resolution 
hydrodynamic simulations.

Further tests may refine our understanding of the environmental scale that is 
important for determining the chemical evolution of dwarf galaxies.  In 
particular, it will be important to examine whether the influence of relatively 
small-scale ($\sim$2 Mpc) environments is more significant to a dwarf galaxy's 
chemical evolution than the larger-scale environment investigated here.  In 
previous work, both \cite{Kreckel15} and \cite{Beygu17} find little evidence to 
support a significant large-scale environmental influence on the gas content, 
chemical content, or star formation rate of void galaxies.  Future work will 
expand on these studies with a larger sample and the possible influence they 
might have on the dwarf galaxies' chemical contents and star formation rates.  
\cite{Beygu17} also investigated any connection between a galaxy's physical 
properties and its location within a void.  We also look to study this possible 
connection with the increased sample size of dwarf galaxies in Douglass \& 
Vogeley (2017, in preparation).

%
%
\acknowledgements
The authors would like to acknowledge Renyue Cen for his help in the 
interpretation of these results.  We would also like to thank the referee, Burcu 
Beygu, for her insightful comments and feedback.

Support for this work was provided by NSF grant AST--1410525.

Funding for the SDSS and SDSS-II has been provided by the Alfred P. Sloan 
Foundation, the Participating Institutions, the National Science Foundation, the 
U.S. Department of Energy, the National Aeronautics and Space Administration, 
the Japanese Monbukagakusho, the Max Planck Society, and the Higher Education 
Funding Council for England.  The SDSS Web Site is \emph{http://www.sdss.org/}.

The SDSS is managed by the Astrophysical Research Consortium for the 
Participating Institutions.  The Participating Institutions are the American 
Museum of Natural History, Astrophysical Institute Potsdam, University of Basil, 
University of Cambridge, Case Western Reserve University, University of Chicago, 
Drexel University, Fermilab, the Institute for Advanced Study, the Japan 
Participation Group, Johns Hopkins University, the Joint Institute for Nuclear 
Astrophysics, the Kavli Institute for Particle Astrophysics and Cosmology, the 
Korean Scientist Group, the Chinese Academy of Sciences (LAMOST), Los Alamos 
National Laboratory, the Max-Planck-Institute for Astronomy (MPIA), the 
Max-Planck-Institute for Astrophysics (MPA), New Mexico State University, Ohio 
State University, University of Pittsburgh, University of Portsmouth, Princeton 
University, the United States Naval Observatory, and the University of 
Washington.


\bibliographystyle{apj}
\bibliography{Doug0127_sources}

\end{document}

%% file: t1_short.tex
\begin{deluxetable}{cccccccccccc}
\tablewidth{0pt}
\tablecolumns{12}
\tablehead{\colhead{Index\tablenotemark{a}} & \colhead{R.A.} & \colhead{Decl.} & \colhead{Redshift} & \colhead{$M_r$} & \multicolumn{2}{c}{$12 + \log \left( \frac{\text{O}}{\text{H}} \right)$} & \multicolumn{2}{c}{$12 + \log \left( \frac{\text{N}}{\text{H}} \right)$} & \multicolumn{2}{c}{$\log \left( \frac{\text{N}}{\text{O}} \right)$} & \colhead{Void/Wall}}
\tablecaption{Analyzed dwarf galaxies\label{tab:Results}}
\startdata
63713 & \RA{09}{20}{04}{.27} & -\dec{00}{30}{08}{.97} & 0.0257 & -16.73 & 7.80 & $\pm$0.41 & 6.83 & $\pm$0.28 & -0.97 & $\pm$0.49 & Wall \\
73537 & \RA{09}{25}{24}{.23} & +\dec{00}{12}{40}{.39} & 0.0250 & -16.94 & 7.94 & $\pm$0.34 & 6.76 & $\pm$0.24 & -1.18 & $\pm$0.41 & Wall \\
75442 & \RA{13}{13}{24}{.25} & +\dec{00}{15}{02}{.95} & 0.0264 & -16.81 & 7.55 & $\pm$0.35 & 6.73 & $\pm$0.24 & -0.82 & $\pm$0.42 & Void \\
168874 & \RA{11}{45}{13}{.16} & -\dec{01}{48}{17}{.68} & 0.0273 & -16.99 & 8.16 & $\pm$0.31 & 6.94 & $\pm$0.21 & -1.21 & $\pm$0.37 & Wall \\
184308 & \RA{09}{39}{09}{.38} & +\dec{00}{59}{04}{.15} & 0.0244 & -16.73 & 7.36 & $\pm$0.43 & 6.71 & $\pm$0.31 & -0.65 & $\pm$0.53 & Wall\\
\enddata
\tablecomments{Five of the 135 dwarf galaxies analyzed from SDSS DR7.  The flux values for all required emission lines can be found in the MPA-JHU value-added catalog.  Metallicity values are calculated using the direct $T_e$ method, with error estimates via a Monte Carlo method.  The void catalog of \cite{Pan12} is used to classify the galaxies as either Void or Wall.  If a galaxy is located too close to the boundary of the SDSS survey to identify whether or not it is inside a void, it is labeled as Uncertain.  Table \ref{tab:Results} is published in its entirety online in a machine-readable format.  A portion is shown here for guidance regarding its form and content.}
\tablenotetext{a}{KIAS-VAGC galaxy index number}
\end{deluxetable}

%% file: t2.tex
\begin{deluxetable}{ccccccccccc}
\tablewidth{0pt}
\tablehead{\colhead{Index\tablenotemark{a}} & \colhead{R.A.} & \colhead{Decl.} & \colhead{Redshift} & \multicolumn{2}{c}{$12 + \log \left( \frac{\text{O}}{\text{H}} \right)$} & \multicolumn{2}{c}{$12 + \log \left( \frac{\text{N}}{\text{H}} \right)$} & \multicolumn{2}{c}{$\log \left( \frac{\text{N}}{\text{O}} \right)$} & \colhead{Void/Wall}}
\tablecaption{Extreme low-metallicity dwarf galaxies\label{tab:lowZ}}
\startdata
268470 & \RA{13}{18}{17}{.82} & +\dec{02}{12}{59}{.83} & 0.0252 & 7.06 & $\pm$0.37 & 6.40 & $\pm$0.25 & -0.66 & $\pm$0.45 & Void \\
1422637 & \RA{14}{18}{12}{.14} & +\dec{13}{59}{33}{.98} & 0.0261 & 7.15 & $\pm$0.41 & 6.29 & $\pm$0.28 & -0.87 & $\pm$0.50 & Wall \\
839665 & \RA{08}{09}{53}{.53} & +\dec{29}{17}{04}{.82} & 0.0281 & 7.18 & $\pm$0.44 & 6.29 & $\pm$0.31 & -0.89 & $\pm$0.54 & Void \\
1168448 & \RA{11}{06}{41}{.00} & +\dec{45}{19}{09}{.28} & 0.0220 & 7.19 & $\pm$0.46 & 6.18 & $\pm$0.32 & -1.01 & $\pm$0.57 & Wall \\
1299291 & \RA{12}{17}{14}{.02} & +\dec{43}{18}{53}{.36} & 0.0233 & 7.21 & $\pm$0.42 & 6.32 & $\pm$0.29 & -0.89 & $\pm$0.51 & Wall \\
1170573 & \RA{11}{05}{39}{.42} & +\dec{46}{03}{28}{.37} & 0.0250 & 7.24 & $\pm$0.34 & 6.16 & $\pm$0.23 & -1.08 & $\pm$0.41 & Wall \\
2288717 & \RA{10}{46}{12}{.18} & +\dec{21}{31}{37}{.37} & 0.0248 & 7.27 & $\pm$0.48 & 6.19 & $\pm$0.33 & -1.08 & $\pm$0.58 & Wall \\
955643 & \RA{11}{42}{03}{.02} & +\dec{49}{21}{25}{.18} & 0.0244 & 7.28 & $\pm$0.44 & 6.12 & $\pm$0.30 & -1.15 & $\pm$0.53 & Wall \\
1344311 & \RA{12}{33}{13}{.64} & +\dec{11}{10}{28}{.46} & 0.0245 & 7.29 & $\pm$0.50 & 6.29 & $\pm$0.34 & -1.00 & $\pm$0.61 & Wall \\
1254352 & \RA{13}{29}{02}{.45} & +\dec{10}{54}{55}{.80} & 0.0237 & 7.30 & $\pm$0.44 & 6.57 & $\pm$0.30 & -0.73 & $\pm$0.54 & Wall \\
1857820 & \RA{08}{45}{00}{.34} & +\dec{27}{16}{47}{.04} & 0.0257 & 7.31 & $\pm$0.48 & 6.48 & $\pm$0.33 & -0.83 & $\pm$0.59 & Wall \\
866876 & \RA{09}{04}{57}{.96} & +\dec{41}{29}{36}{.42} & 0.0240 & 7.32 & $\pm$0.40 & 6.43 & $\pm$0.28 & -0.89 & $\pm$0.49 & Wall \\
833588 & \RA{08}{43}{10}{.71} & +\dec{43}{08}{53}{.58} & 0.0245 & 7.34 & $\pm$0.41 & 6.14 & $\pm$0.29 & -1.21 & $\pm$0.50 & Wall \\
283263 & \RA{14}{14}{12}{.88} & +\dec{01}{50}{12}{.88} & 0.0255 & 7.35 & $\pm$0.43 & 6.45 & $\pm$0.29 & -0.90 & $\pm$0.52 & Wall \\
184308 & \RA{09}{39}{09}{.38} & +\dec{00}{59}{04}{.15} & 0.0244 & 7.36 & $\pm$0.43 & 6.71 & $\pm$0.31 & -0.65 & $\pm$0.53 & Wall \\
1389829 & \RA{14}{31}{01}{.38} & +\dec{38}{04}{21}{.50} & 0.0269 & 7.41 & $\pm$0.46 & 6.50 & $\pm$0.32 & -0.91 & $\pm$0.56 & Wall \\
858951 & \RA{09}{31}{39}{.60} & +\dec{49}{49}{56}{.85} & 0.0251 & 7.46 & $\pm$0.46 & 6.33 & $\pm$0.32 & -1.13 & $\pm$0.57 & Wall \\
1270221 & \RA{13}{27}{39}{.85} & +\dec{50}{54}{09}{.69} & 0.0295 & 7.49 & $\pm$0.43 & 6.45 & $\pm$0.30 & -1.04 & $\pm$0.52 & Wall \\
431383 & \RA{08}{58}{44}{.96} & +\dec{50}{29}{58}{.98} & 0.0230 & 7.52 & $\pm$0.60 & 6.26 & $\pm$0.31 & -1.26 & $\pm$0.68 & Void \\
75442 & \RA{13}{13}{24}{.25} & +\dec{00}{15}{02}{.95} & 0.0264 & 7.55 & $\pm$0.35 & 6.73 & $\pm$0.24 & -0.82 & $\pm$0.42 & Void \\
1322765 & \RA{14}{15}{05}{.58} & +\dec{36}{22}{57}{.77} & 0.0273 & 7.57 & $\pm$0.40 & 6.64 & $\pm$0.28 & -0.93 & $\pm$0.48 & Wall\\
\enddata
\tablecomments{Details of 21 extreme low gas-phase metallicity ($12 + \log(\text{O}/\text{H}) < 7.6$) galaxies identified in Paper 1.  While the N/H values for all these galaxies are also extremely low, the N/O ratios span the range covered by the remainder of the dwarf galaxy sample studied.  Further study of these galaxies is recommended to confirm the abundance values and identify any shared characteristics.}
\tablenotetext{a}{KIAS-VAGC galaxy index number}
\end{deluxetable}